\documentclass[lettersize,journal]{IEEEtran}
\usepackage{amsmath,amsfonts}
\usepackage{algorithmic}
\usepackage{algorithm}
\usepackage{array}
\UseRawInputEncoding
\usepackage{textcomp}
\usepackage{stfloats}
\usepackage{url}
\usepackage{verbatim}
\usepackage{graphicx}
\usepackage{amssymb} 
\usepackage{amsfonts}
\usepackage{amsmath} 
\usepackage{physics} 
\usepackage{algorithmic}  
\usepackage{algorithm} 
\usepackage{graphicx}
\usepackage{color} 
\usepackage{bm}
\usepackage{epstopdf} 
\usepackage{subfigure}
\usepackage{booktabs}  
\usepackage[dvipsnames]{xcolor} 
\usepackage{cite} 
\usepackage{balance}
\usepackage{setspace}  
\usepackage{amsthm}
\usepackage{amsmath} 
\usepackage{hyperref}
\usepackage{balance}
\usepackage{float}

\newtheoremstyle{note}
{3pt}%
{3pt}%
{}%
{\parindent}%
{\rmfamily \bfseries}%
{:}%
{5pt}%
{}%

\newtheorem{ppn}{Proposition}
 
\newtheorem{remark}{Remark}

\newtheorem*{pf}{Proof}  

\newcommand{\ppnref}[1]{\textbf{Proposition \ref{#1}}}

\newcommand{\algref}[1]{\textbf{Algorithm \ref{#1}}}

\newcommand{\figref}[1]{Fig. \ref{#1}}
\newcommand{\secref}[1]{Section \ref{#1}}

\begin{document}
	
	\title{Interference in Spectrum-Sharing Integrated Terrestrial and Satellite Networks: Modeling, Approximation, and Robust Transmit Beamforming}
	
	\author{Wenjing Cao, Yafei Wang, \textit{Graduate Student Member},
    \textit{IEEE},
    Tianxiang Ji, Tianyang Cao, \\ Wenjin Wang, \textit{Member}, \textit{IEEE},
   Symeon Chatzinotas, \textit{Fellow}, \textit{IEEE}, Bj{\"o}rn Ottersten, \textit{Fellow}, \textit{IEEE}
		\thanks{Manuscript received xxx; revised xxx.  \textit{}}
		\thanks{Wenjing Cao, Yafei Wang, and Wenjin Wang are with the National Mobile Communications Research Laboratory, Southeast University, Nanjing 210096, China, and also with Purple Mountain Laboratories, Nanjing 211100, China (e-mail: caowj@seu.edu.cn; wangyf@seu.edu.cn; wangwj@seu.edu.cn).}
        \thanks{Tianxiang Ji and Tianyang Cao are with China Mobile Group Design Institute Company Ltd., Beijing 100089, China (e-mail: jitianxiang@cmdi.chinamobile.com; caotianyang@cmdi.chinamobile.com).}
		\thanks{Symeon Chatzinotas and Bj{\"o}rn Ottersten are with the Interdisciplinary Centre for Security, Reliability and Trust (SnT), University of Luxembourg, Luxembourg (e-mail: symeon.chatzinotas@uni.lu;  bjorn.ottersten@uni.lu).}}
	
	\markboth{}%
	{Shell \MakeLowercase{\textit{et al.}}: A Sample Article Using IEEEtran.cls for IEEE Journals}
	
	
	\maketitle

	\begin{abstract}     
     \textcolor{black}{This paper investigates robust transmit (TX) beamforming from the satellite to user terminals (UTs), based on statistical channel state information (CSI).  The proposed design specifically targets the mitigation of satellite-to-terrestrial interference in spectrum-sharing integrated terrestrial and satellite networks. 
     By leveraging the distribution information of terrestrial UTs, we first establish an interference model from the satellite to terrestrial systems without shared CSI.
  Based on this, robust TX beamforming schemes are developed under both the interference threshold and the power budget. Two optimization criteria are considered: satellite weighted sum rate maximization and mean square error minimization. The former achieves a superior achievable rate performance through an iterative optimization framework, whereas the latter enables a low-complexity closed-form solution at the expense of reduced rate, with interference constraints satisfied via a bisection method.}
  To avoid complex integral calculations and the dependence on user distribution information in inter-system interference evaluations, we propose a terrestrial base station position-aided approximation method, and the approximation errors are subsequently analyzed.
   Numerical simulations validate the effectiveness of our proposed schemes.
	\end{abstract}
	

	%
	\IEEEpeerreviewmaketitle

\section{Introduction}
\IEEEPARstart{S}{atellite} communications are crucial for enabling ubiquitous connectivity in the 6th generation (6G), playing a significant role in achieving broad coverage across space, air, ground, and sea \cite{8088533,8746876}. However, the competition for spectrum among various wireless communication systems has intensified, especially in frequency bands below 6 GHz \cite{8403528,5586925,7811844}, posing significant challenges to the allocation of dedicated frequencies for satellite communications.  \textcolor{black}{Spectrum sharing in integrated terrestrial and satellite networks (ITSNs) has emerged as a promising solution \cite{9052737,9385374,9485040,10645749,10301688,9749196,9344715}, allowing satellites and terrestrial communication networks to utilize the same spectrum to enhance spectrum efficiency, as demonstrated by early efforts such as cognitive radio for satellite communications (CoRaSat) technology \cite{7336495,lagunas2015power} and shared access terrestrial-satellite backhaul network enabled by smart antennas (SANSA) project \cite{sat1207,lagunas2017carrier}.} However, the overlap of the operating spectrum between terrestrial and satellite systems \textcolor{black}{may} cause inter-system interference within their overlapping coverage areas, leading to degradation in overall system performance \cite{7986310,7559751,7387774}. \textcolor{black}{In sub-6GHz ITSN scenarios, where satellite communications is considered complementary to terrestrial mobile networks, satellite user terminals (UTs) are often located outside the coverage of terrestrial base stations (BSs), making interference from BSs to satellite UTs negligible \cite{9110855,Pastukh2023ChallengesOU}.} Nevertheless, due to the broader coverage of satellite beams \cite{10179219}, \textcolor{black}{interference from satellites to terrestrial UTs} may still occur at the interface of the two systems. \textcolor{black}{Therefore, mitigating interference between terrestrial and satellite systems is essential for efficient spectrum sharing of ITSNs.}

\textcolor{black}{Transmit (TX) beamforming} is widely recognized as an effective technique for interference management in satellite communication systems \cite{7887756,10256078}. \textcolor{black}{Studies in CoRaSat and SANSA primarily focused on high-frequency bands (e.g., Ka and Ku bands) and geostationary earth orbit satellites, while recent developments have shifted interest toward spectrum sharing in sub-6GHz bands and low earth orbit (LEO) satellites \cite{9992172}, driven by direct-to-cell satellite initiatives such as AST SpaceMobile's BlueWalker 3 and SpaceX's Starlink V2 Mini \cite{10820534}. Benefiting from their extremely large antenna arrays onboard, these satellites enable advanced beamforming capabilities \cite{10068542}.}
\textcolor{black}{Conventional TX beamforming methods,} such as maximum ratio transmission (MRT) beamforming, zero-forcing (ZF) beamforming, minimum mean square error (MMSE) beamforming, weighted MMSE (WMMSE) beamforming, total power minimization beamforming, and energy efficiency beamforming \cite{1494410,1391204,4712693,wang2024,10437228,6832894}, require perfect instantaneous channel state information (iCSI) to achieve optimal performance. However, due to the long satellite-to-terrestrial distance and the high mobility of satellites, obtaining accurate iCSI is challenging, and beamforming design based on statistical CSI (sCSI), e.g., the angle of departure (AoD), becomes a more practical approach. For example, a beamforming method based on sCSI, including angle and power information, was proposed in \cite{9110855} with the objective of maximizing the average signal-to-interference-plus-noise ratio (SINR).
To against errors in the AoD within sCSI, \cite{9815078} proposed a robust beamforming scheme with a more practical per-antenna power constraint.

Although the aforementioned \textcolor{black}{TX} beamforming schemes effectively mitigate intra-system user interference, they cannot be directly applied to ITSN systems, as they do not consider inter-system interference. \textcolor{black}{To address this challenge, some works have proposed dedicated \textcolor{black}{TX} beamforming designs to handle inter-system interference.} To eliminate interference from terrestrial BSs to satellite UTs, \cite{10038746} proposed a \textcolor{black}{TX} beamforming method at the terrestrial BS side under the minimum user rate constraint, and this terrestrial interference was further restricted to an interference power threshold in \cite{10.1049/iet-com.2018.5313,6636830}. Joint \textcolor{black}{TX} beamforming design for terrestrial and non-terrestrial systems based on shared CSI is another research avenue for interference management. \cite{9485040,9502017} proposed joint \textcolor{black}{TX} beamforming designs subject to minimum user rate constraints, while \cite{8886590} optimized joint beamforming to meet the minimum SINR for each UT in ITSNs. \textcolor{black}{In sub-6GHz ITSNs, where the terrestrial system serves as the primary network and interference from BSs to satellite UTs is negligible \cite{9110855,Pastukh2023ChallengesOU}, mitigating satellite-to-terrestrial interference becomes a more practical and relevant concern.} To reduce satellite interference, \cite{8933099} proposed a \textcolor{black}{TX} beamforming design with cloud-based CSI and introduced an interference threshold for terrestrial UTs, with the objective of maximizing the satellite UT rate. In \cite{8894065}, the \textcolor{black}{TX} beamforming leveraged shared CSI to mitigate satellite interference while constraining a minimum rate for each terrestrial UT.
Compared to joint satellite-terrestrial beamforming designs, optimizing \textcolor{black}{TX} beamforming solely on the satellite side significantly reduces complexity due to the lower optimization dimensions.

\textcolor{black}{Most existing \textcolor{black}{TX} beamforming schemes for ITSNs rely on CSI sharing between the satellite and the terrestrial system \cite{8894065,8933099,8292945,8886590,8353853}, which incurs substantial additional communication overhead and suffers from CSI delay due to mobility.}
However, CSI sharing in ITSNs poses significant challenges for current system architectures due to the lack of protocol support for shared CSI transmission and the large volume of real-time CSI data from terrestrial UTs, which requires collaboration between terrestrial and satellite operators\cite{20251017998213}.
Thus, a key question arises: \textit{How to design robust \textcolor{black}{TX} beamforming without shared CSI against satellite-to-terrestrial interference in ITSNs?} This is the central issue to be investigated in this paper. In summary, our principal contributions are as follows:

\begin{itemize}  
\item 
\textcolor{black}{To address the challenge of CSI sharing in ITSNs,} we develop an integral-form interference model free of shared CSI for the satellite downlink. This model targets the interference from the satellite to terrestrial UTs. \textcolor{black}{Specifically, we derive the interference power as an integral expression, utilizing prior knowledge of the corresponding BS positions and the distribution of terrestrial users.  Thus, this interference model eliminates the dependence on CSI sharing in ITSNs.}
    
\item Based on the proposed interference model and statistical CSI, we \textcolor{black}{design} robust satellite \textcolor{black}{TX} beamforming schemes for interference mitigation in ITSNs. The weighted sum rate (WSR) maximization problem for the satellite is formulated under the interference threshold and power budget, and solved iteratively by transforming the fractional objective into a convex form using multidimensional complex quadratic transformation (MCQT). To reduce complexity, we reformulate the problem as an equivalent WMMSE problem and derive the closed-form beamforming matrix for iteration from the Karush-Kuhn-Tucker (KKT) conditions.

\item \textcolor{black}{To further simplify the design, we propose a low-complexity and robust satellite TX beamforming scheme. We derive the MMSE problem for the satellite under the interference and power constraints, obtain the optimal closed-form solution, and apply a bisection method to satisfy the interference threshold for terrestrial UTs. This scheme avoids high-dimensional iterative updates, offering a favorable trade-off between performance and computational efficiency.}
	
\item  
We propose an interference approximation scheme that uses terrestrial BS position information to approximate the integral-form interference term. This scheme reduces the complexity of interference integral calculations and the challenge of obtaining real-time user distributions. Furthermore, we analyze the key factors affecting the approximation error, providing guidance for the utilization of this approximation. 
\end{itemize}
	
The rest of the article is organized as follows: The system model is established in \secref{System}. The WSR beamforming problem and the corresponding schemes are investigated in \secref{Algorithm-BC}. The MMSE beamforming problem and its closed-form solution are derived in \secref{Algorithm-A}. In \secref{Approximation}, we propose an interference approximation scheme and analyze the approximation error. Simulation results are provided in \secref{Simulation}, and \secref{Conclusion} concludes this article.

{\textit{Notation}}: $x, {\bf x}, {\bf X}$ represent scalar, column vector, matrix. $(\cdot)^T$, $(\cdot)^{*}$, $(\cdot)^H$, and $(\cdot)^{-1}$ respectively denote the transpose, conjugate, transpose-conjugate, and inverse operations. $\left \|\cdot\right \|_{2}$ denotes $l_2$-norm. $\left \|\cdot\right \|_{F}^2$ denotes frobenius norm. $\otimes$ is the Kronecker product operation. The operator ${\rm Tr}\left\{\cdot\right\}$ represents the matrix trace. $\mathbb{E}\left\{\cdot\right\}$ denotes the expectation. ${{\rm diag}\{{\bf a}\}}$ represents the diagonal matrix whose diagonal elements are composed of ${\bf a}$. $[{\bf X}]_{i,j}$ denotes the $(i,j)$-th element of ${\bf x}$. The expression $\mathcal{C}\mathcal{N}(\mu, \sigma^2)$ denotes circularly symmetric Gaussian distribution with expectation $\mu$ and variance $\sigma^2$. ${\mathbb{R}}^{M\times N}$ and ${\mathbb{C}}^{M\times N}$ represent the set of $M\times N$ dimension real- and complex-valued matrixes. $\nabla f$ denotes gradient of function $f(\cdot)$. $k\in \mathcal{K}$ means element $k$ belongs to set $\mathcal{K}$.

\section{System Model}\label{System}
\begin{figure}[tp]
	\centering
\includegraphics[width=0.97\linewidth,trim=0.5cm 2cm 0.5cm 0.2cm,clip]{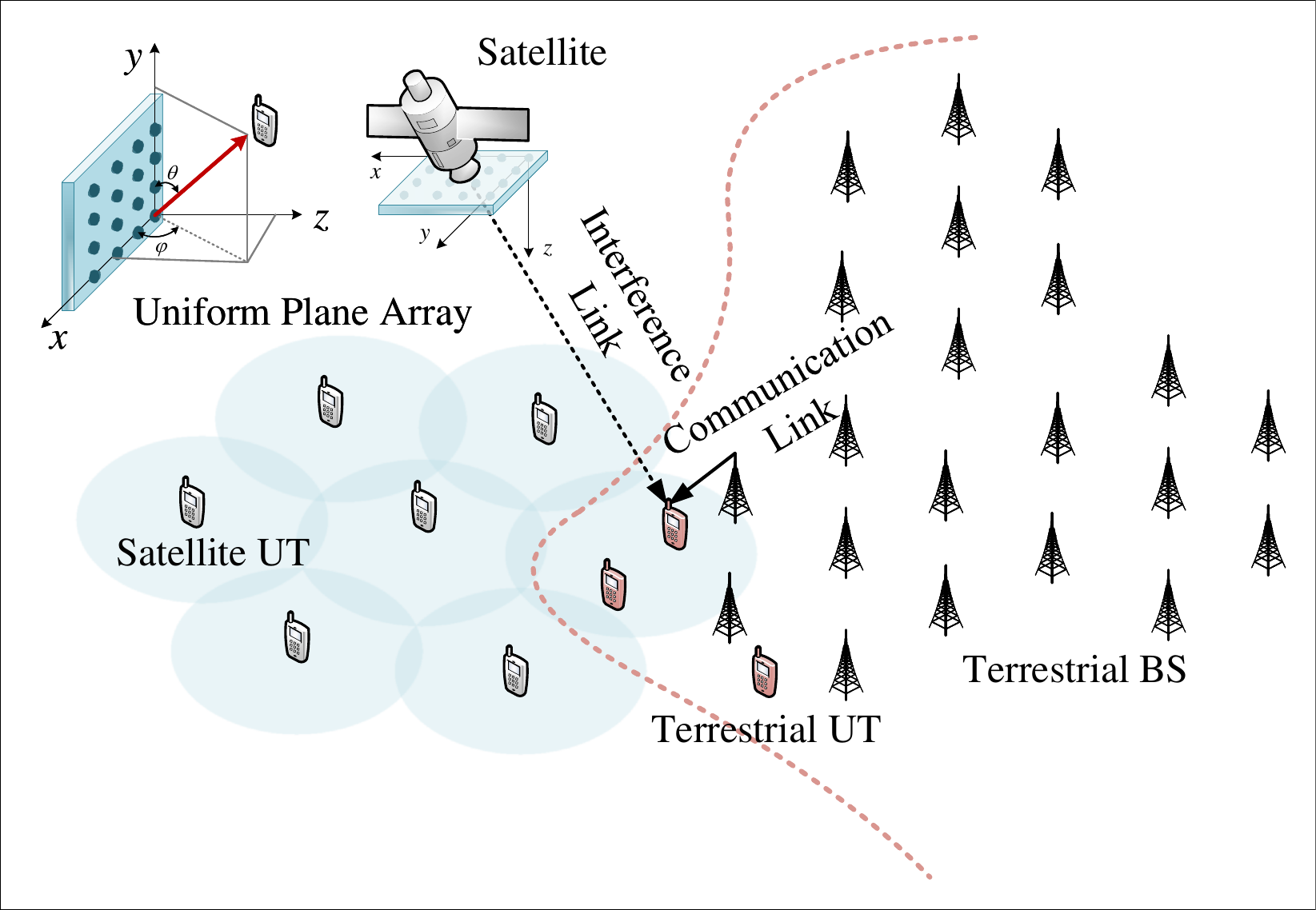}
	\caption{The system architecture of the ITSN.}
	\label{scenario}
\end{figure}

	In \figref{scenario}, we consider a spectrum-sharing \textcolor{black}{satellite downlink} scenario in ITSNs, \textcolor{black}{where the primary terrestrial network coexists with a secondary satellite network.} \textcolor{black}{A LEO satellite serves $K_{\rm S}$ satellite UTs, and \(N_\mathrm{G}\) terrestrial BSs serve $K_{\rm G}$ terrestrial UTs. The satellite and terrestrial systems have partially overlapping coverage areas at the edges \cite{20251017998213}, within which a portion of terrestrial UTs reside.} Both the terrestrial and satellite communication systems share the sub-6GHz spectrum for downlink transmission in response to the requirements of the 3rd Generation
 Partnership Project (3GPP) \cite{10301688,3GPPTR38.811}.

\vspace{-1mm}
\subsection{Channel and Signal Model}\label{channel and signal model}
\vspace{-1mm}

	We consider a satellite equipped with a large-scale uniform planar array (UPA) composed of \( M_\mathrm{S} = M_x \times M_y \) antennas, where \( M_x \) and \( M_y \) denote the number of antenna elements along the \( x \)-axis and \( y \)-axis, respectively. \textcolor{black}{Adopting a ray-tracing-based channel modeling approach,} the space-domain channel vector  $\mathbf{h}_{\mathrm{ss},k}\left(t,f\right)\in \mathbb{C}^{M_\mathrm{S} \times 1}$ for the satellite UT \( k \) at time instant \( t \) and frequency \( f \) can be modeled as \cite{9110855,iet12629,9427230}
	\begin{align}
\mathbf{h}_{\mathrm{ss},k}\left(t,f\right)=\mathrm{e}^{j2 \pi\left(t v_k^{\rm{sat}}- f \tau_{k}^{\rm{min}}\right)} g_{k}^{\mathrm{ss}}\left(t,f\right)  \mathbf{v}_k^{\rm{ss}},
    \label{1}
    \end{align}
where \(g_{k}^{\mathrm{ss}} \) represents the LEO satellite downlink channel gain. Due to the line-of-sight (LoS) propagation characteristics of satellite communication, it is assumed that the channel gain $g_{k}^{\mathrm{ss}}\left(t,f\right)$ \textcolor{black}{follows a Rician distribution} with the Rician factor $\kappa_{k}$ and power $\mathbb{E}\left\{|g_{k}^{\mathrm{ss}}(t, f)|^{2}\right\}=(\gamma_{k}^{\rm{ss}})^2$.
Specifically, the real and imaginary parts of \(g_{k}^{\mathrm{ss}}\) are independently and identically real-valued Gaussian distributed with mean  $\gamma_{k}^{\rm{ss}}\sqrt{\frac{\kappa_{k}}{2(\kappa_{k }+1)}}$ and variance  $\frac{(\gamma_{k}^{\rm{ss}})^2}{2(\kappa_{k}+1)}$, respectively \cite{4655459,9815078}. Additionally, $v_k^{\rm{sat}}$ refers to the Doppler shift caused by the motion of the LEO satellite, and \(\tau_{k}^{\rm{min}}\) refers to the minimum propagation delay of the $k$-th satellite UT. \textcolor{black}{The UPA TX characteristics can be represented by the vector \(\mathbf{v}_k^{\rm{ss}} = \mathbf{v}_k^{\rm{ss}}\left(\vartheta_{k}^{x}\right) \otimes \mathbf{v}_k^{\rm{ss}}\left(\vartheta_{k}^{y}\right)\).}
As shown in \figref{scenario}, the channel space angles \( \vartheta_{k}^{x}=\sin \theta_{k} \cos \varphi_{k} \) and \( \vartheta_{k}^{y}=\cos \theta_{k} \) represent the direction cosines of the UT with respect to the UPA along the \( x \)-axis and \( y \)-axis, respectively. Herein, \( \theta_{k} \) denotes the elevation angle and \( \varphi_{k} \) denotes the azimuth angle from the satellite to the \( k \)-th satellite UT.
$\mathbf{v}_k^{\rm{ss}}\left(\vartheta_{k}^{x}\right)$ and   $\mathbf{v}_k^{\rm{ss}}\left(\vartheta_{k}^{y}\right) $ are the steering vectors in the \( x \)-axis and \( y \)-axis as follows
\begin{align}
\!\mathbf{v}_k^{\rm{ss}}\!\left(\vartheta_{k}^{x}\right)\!=\! \frac{1}{\sqrt{M_{x}}}\!\left[1, \mathrm{e}^{-\frac{j 2 \pi}{\lambda} d_{x} \vartheta_{k}^{x}},\ldots,   \mathrm{e}^{- \frac{j 2 \pi}{\lambda} d_{x}\left(M_{x}-1\right) \vartheta_{k}^{x}}\right]^{T}\!\!\!,\label{2}\\
\!\mathbf{v}_k^{\rm{ss}}\!\left(\vartheta_{k}^{y}\right)\! =\!\frac{1}{\sqrt{M_{y}}}\!\left[1, \mathrm{e}^{-\frac{j 2 \pi}{\lambda} d_{y} \vartheta_{k}^{y}} ,\ldots,\mathrm{e}^{-\frac{j 2 \pi}{\lambda} d_{y}\left(M_{y}-1\right) \vartheta_{k}^{y}}\right]^{T}\!\!\!,\label{3}
\end{align}
where \( d_x = d_y = \frac{\lambda}{2} \) represent the distances between adjacent antenna elements along the \( x \)-axis and \( y \)-axis.

In ITSNs, the satellite and \( N_\mathrm{G} \) terrestrial BSs share the spectrum to provide \textcolor{black}{service} to corresponding UTs. \textcolor{black}{Since satellites typically complement terrestrial mobile communication, we assume that in the sub-6GHz ITSN scenario, the satellite switches off the coverage over areas with terrestrial connectivity and satellite UTs are usually located outside the coverage area of terrestrial BSs, making terrestrial interference to satellite UTs negligible \cite{8933099,Pastukh2023ChallengesOU}.} The signal \( y_{\mathrm{s},k} \) received by the \( k \)-th satellite UT can be expressed as
	\begin{align}
    \textstyle
        y_{\mathrm s,k}= \mathbf{h}_{\mathrm{ss},k}^H \sum_{i=1}^{K_\mathrm{S}}\mathbf{p}_{\mathrm s,i}x_{\mathrm s,i}+n_{\mathrm s,k},
		\label{4}
	\end{align}
herein, \( \mathbf{p}_{\mathrm{s},i} \in \mathbb{C}^{M_\mathrm{S} \times 1} \) is the beamforming vector, \( x_{\mathrm{s},i} \) is the symbol sent by the satellite to the \( i \)-th satellite UT and \( n_{\mathrm{s},k} \) denotes the additive noise following $\mathcal{CN}(0, \sigma_k^{2})$.
For brevity, we combine the received signals of all satellite UTs and reformulate the signal model as
	\begin{align}
		\begin{split}
			\mathbf{y}_{\mathrm s}=\mathbf{H}_\mathrm{ss}^H\mathbf{P}\mathbf{x}_\mathrm{s}+\mathbf{n}_\mathrm{s} \in \mathbb{C}^{K_\mathrm{S} \times 1},
		\end{split}
		\label{5}
	\end{align}
where channel matrix \( \mathbf{H}_{\mathrm{ss}}=\left[\mathbf{h}_{\mathrm{ss}, 1}, \ldots, \mathbf{h}_{\mathrm{ss}, K_{\mathrm{S}}}\right] \in \mathbb{C}^{M_{\mathrm{S}} \times K_{\mathrm{S}}} \), beamforming matrix \( \mathbf{P} = [{\mathbf{p}_{\mathrm{s,1}}, \ldots, \mathbf{p}_{\mathrm{s},K_\mathrm{S}}}] \in \mathbb{C}^{M_\mathrm{S} \times K_\mathrm{S}} \). Besides, \( \mathbf{x}_\mathrm{s} = {[x_{\mathrm{s,1}}, \ldots, x_{\mathrm{s},K_\mathrm{S}}]}^T \in \mathbb{C}^{K_\mathrm{S} \times 1} \) is the matrix of symbols sent by the satellite satisfying \( \mathbb{E}\left\{\mathbf{x}_\mathrm{s} \mathbf{x}_\mathrm{s}^{H}\right\} = \mathbf{I} \).

Despite the benefits of spectrum sharing between satellite and terrestrial networks, inter-system interference could degrade system performance. Effective interference mitigation, particularly in overlapping coverage regions of ITSNs, is essential for successful spectrum sharing \cite{10301688}. To simplify the analysis, we neglect the interference among terrestrial BSs and focus on the interference from the satellite to terrestrial UTs. The average interference power to each terrestrial UT can be expressed as
\begin{align}
	\begin{split}
I_{\rm avg}&=\mathbb{E}_{\mathbf{x}_{\mathrm{s}}}\left\{\frac{1}{K_{\rm G}} \operatorname{Tr}\left\{\mathbf{H}_{\mathrm{sg}}^H\mathbf{P}\mathbf{x}_{\mathrm{s}}\mathbf{x}_{\mathrm{s}}^H\mathbf{P}^H\mathbf{H}_{\mathrm{sg}}\right\}\right\}\\
&=\frac{1}{K_{\rm G}} \operatorname{Tr}\left\{\mathbf{P}^H \mathbf{H}_{\mathrm{sg}}\mathbf{H}_{\mathrm{sg}}^H\mathbf{P}\right\},
\end{split}
\label{8}
\end{align}
where \( \mathbf{H}_\mathrm{sg}  \in \mathbb{C}^{M_{\mathrm{S}} \times K_\mathrm{G}} \) denotes the satellite-to-terrestrial UT channel matrix {based on iCSI}. \textcolor{black}{This interference channel is modeled as a Rician fading channel with LoS propagation, consistent with the satellite signal link, since LoS terrestrial UTs typically suffer stronger satellite interference than non-LoS ones. The interference power, governed by power flux density limits that adapt to the beam footprint, must remain below \( I_{\rm thr} \) in accordance with 3GPP and satellite operator standards \cite{3GPPTR38.863,fcc_sat_loa}.} 

\subsection{Foundation of Beamforming in ITSNs: sCSI or CSI-Free?}\label{robust transmission}

Benefiting from the large-scale antenna arrays onboard the satellite, we leverage the \textcolor{black}{TX beamforming technique} to mitigate the \textcolor{black}{interference from the satellite to terrestrial UTs.} While conventional \textcolor{black}{TX} beamforming design typically relies on iCSI availability, acquiring precise iCSI is generally infeasible for the satellite due to the long propagation delay, large Doppler shift, and limited pilot overhead \cite{8894065,8795582}. Furthermore, the frequent updating of \textcolor{black}{TX} beamforming vectors based on iCSI of $\mathbf{H}_{\rm ss}$ and $\mathbf{H}_{\rm sg}$ presents significant challenges for satellite payload implementation. Consequently, we design robust \textcolor{black}{TX} beamforming utilizing the following information in ITSNs:
 \subsubsection{sCSI for Intra-Satellite System Channels} In the satellite system, sCSI for satellite UTs includes
\begin{align}
&\bar \gamma_{k}^{\rm{ss}}\triangleq \mathbb{E}\left\{g_k^{\rm ss}(t,f)\right\}=\gamma_{k}^{\rm{ss}}\sqrt{\frac{\kappa_{k}}{2(\kappa_{k }+1)}}(1+j),\\
&\textstyle\mathbb{E}\left\{\mathbf{H}_{\mathrm{ss}} \mathbf{H}_{\mathrm{ss}}^{H}\right\}=\sum_{k=1}^{K_{\rm S}}(\gamma_{k}^{\rm{ss}})^2\mathbf{v}_{k}^{\rm{ss}}{\mathbf{v}_{k}^{\rm{ss}}}^H,
\end{align}
which are assumed to vary slowly in satellite systems \cite{9110855}.
\subsubsection{CSI-Free Interference Model for Inter-Satellite and Terrestrial Systems} The average interference power can be further expressed in statistical form
\begin{align}
	\begin{split}
I_{\rm avg}&=\mathbb{E}_{\mathbf{H}_{\mathrm{sg}},\mathbf{x}_{\mathrm{s}}}\left\{\frac{1}{K_{\rm G}} \operatorname{Tr}\left\{\mathbf{H}_{\mathrm{sg}}^H\mathbf{P}\mathbf{x}_{\mathrm{s}}\mathbf{x}_{\mathrm{s}}^H\mathbf{P}^H\mathbf{H}_{\mathrm{sg}}\right\}\right\} \\
&=\frac{1}{K_{\rm G}} \operatorname{Tr}\left\{\mathbf{P}^H \mathbb{E}\left\{\mathbf{H}_{\mathrm{sg}}\mathbf{H}_{\mathrm{sg}}^H\right\}\mathbf{P}\right\}\\&=\frac{1}{K_{\rm G}} \operatorname{Tr}\left\{\mathbf{P}^H\boldsymbol{\Upsilon} _{\mathrm{sg}}\mathbf{P}\right\},
\end{split}
\label{8_1}
\end{align}
 we define \(\boldsymbol{\Upsilon}_{\mathrm{sg}} =\mathbb{E}\left\{\mathbf{H}_{\mathrm{sg}}\mathbf{H}_{\mathrm{sg}}^H\right\}=\sum_{k=1}^{K_{\rm G}}(\gamma_{k}^{\rm{sg}})^2\mathbf{v}_{k}^{\mathrm{sg}}{\mathbf{v}_{k}^{\mathrm{sg}}}^H\). \textcolor{black}{Due to the numerous interfered terrestrial UTs, obtaining all of their sCSI may still incur significant.  Moreover, the satellite array exhibits limited spatial resolution, making it difficult to discriminate among closely located terrestrial UTs. To address these issues,  we convert the sum of the discrete terrestrial UT channels into an integral of the user distribution over the terrestrial BS coverage.} The integral is given by
\begin{equation}
    \begin{aligned}
&\boldsymbol{\Upsilon}_{\mathrm{sg}}  =\mathbb{E}\left\{\mathbf{H}_{\mathrm{sg}}
\mathbf{H}_{\mathrm{sg}}^H\right\}=\sum_{n=1}^{N_{\mathrm{G}}} \mathbb{E}\left\{\sum_{k=1}^{\bar{K}_{\mathrm{G}}} \left|g_{n, k}^{\mathrm{sg}}(t,f)\right|^{2}\mathbf{v}_{n, k}^{\mathrm{sg}} {\mathbf{v}_{n, k}^{\mathrm{sg}}}^H\right\}\\
&\!=\!\!\!\sum_{n=1}^{N_{\mathrm{G}}} \!\int_0^{2\pi}\!\!\!\!\!\int_0^{R_{\rm bs}}\!\!\frac{c^2M_x M_y G_{\rm T}G_{\rm R}\!\mathbf{V}_n^{\rm sg}(r_n,\!\phi_n) \!f(r_n,\!\phi_n)r_n{\rm d}r_n{\rm d}\phi_n}{\left(4\pi f d_{n}(r_n,\phi_n)\right)^2}
,
\end{aligned}
\end{equation}
where $N_{\mathrm{G}}$ and $\bar{K}_{\mathrm{G}}$ are the numbers of terrestrial BSs and UTs per BS, respectively. We have \( \mathbb{E}\left\{|g_{n, k}^{\mathrm{sg}}(t, f)|^{2}\right\} = (\gamma_{n,k}^{\rm{sg}})^{2} = \frac{M_x M_yG_{\rm T} G_{\rm R} c^2}{\left(4\pi f d_{n,k}\right)^2} \) and $\mathbf{V}_n^{\rm sg}(r_n,\phi_n)=\mathbf{v}_n^{\rm sg}(r_n,\phi_n)\mathbf{v}_n^{\rm sg}(r_n,\phi_n)^H$. \( G_{\rm T} \) and \( G_{\rm R} \) are the per-antenna transmit gain and receive gain, respectively. $f(r_n,\phi_n)$ is the probability density function of user distribution. The propagation distance between the satellite and the UT at the \(n\)-th terrestrial BS, given the UT's polar coordinates \( \phi_n \) and \( r_n \) representing the angular and radial positions, respectively, can be computed by
\begin{align}
	\begin{split}
d_n(r_n,\phi_n)=\sqrt{h_{\rm{sat}}^{2}+R_n^2  + r_n^2+ 2R_nr_n\cos( \psi_n- \phi_n)},
\end{split}
\label{}
\end{align}
where \( h_{\rm{sat}} \) denotes the satellite height, \( \psi_n \) the polar angle of  the \( n \)-th terrestrial BS, and \( R_n \) the distance from the \( n \)-th terrestrial BS to the sub-satellite point. Then we compute the $(i,j)$-th element of $\mathbf{V}_n^{\rm sg}(r_n,\phi_n)$ as
\begin{align}
	\begin{split}
\left[\mathbf{V}_n^{\rm sg}(r_n,\phi_n)\right]_{i,j}=\frac{\mathrm{e}^{j\pi[\left(m_{a}-m_{p}\right) \vartheta_{n}^{x}+\left(n_{b}-n_{q}\right) \vartheta_{n}^{y}]}}{M_xM_y}.
\end{split}
\end{align}
The element indices are determined by the expressions \(i = n_q M_x + m_p + 1\) and \(j = n_b M_x + m_a + 1\), where the indices follow the ordering: \( n_q = 0, \dots, M_y - 1\), \( m_p = 0, \dots, M_x - 1\), \( n_b = 0, \dots, M_y - 1\) and \( m_a = 0, \dots, M_x - 1\). The channel space angles are computed as \(\vartheta_{n}^{x}(r_n,\phi_n) = \frac{R_n \cos \psi_n + r_n \cos \phi_n}{R_{\rm sat}}\) and \(\vartheta_{n}^{y}(r_n,\phi_n) = \frac{R_n \sin \psi_n + r_n \sin \phi_n}{R_{\rm sat}}\) where \( R_{\rm{sat}} \) is the satellite coverage radius. To simplify,  we replace $\varpi_n(r_n,\phi_n) =(m_a-m_p)\vartheta_{n}^{x}+(n_b-n_q)\vartheta_{n}^{y}$. Thus, we obtain the integral form of the interference channel term \( \boldsymbol{\Upsilon}^{\mathrm{int}}_{\mathrm{sg}} \), and its \((i,j)\)-th element can be rewritten as
\begin{align}
	\begin{split}
\left[\boldsymbol{\Upsilon}_{\mathrm{sg}}^{\rm int}\right]_{i, j}=\sum_{n=1}^{N_{\rm G}}
\int_{0}^{2\pi}\int_{0}^{R_{\mathrm{bs}}}&\frac{{G_{\rm T}G_{\rm R} c^2} \mathrm{e}^{j\pi\varpi_n(r_n,\phi_n)}}{{ \left(4\pi fd_n(r_n,\phi_n)\right)}^2}\\& \times f(r_n,\phi_n)r_n\mathrm{d}r_n\mathrm{d}\phi_n.
\end{split}
\label{integral}
\end{align}
This integral-form interference model, free of shared CSI from terrestrial UTs, leverages BS positions and terrestrial user distribution, thereby reducing the need for extensive pilot overhead, latency, and computational complexity.

In Sections \ref{Algorithm-BC} and \ref{Algorithm-A}, we use the integral-form interference channel term \( \boldsymbol{\Upsilon}^{\mathrm{int}}_{\mathrm{sg}} \), which is free of shared CSI, along with sCSI, including $\mathbb{E}\left\{g_k^{\rm ss}(t,f)\right\}$ and $\mathbb{E}\left\{\mathbf{H}_{\mathrm{ss}} \mathbf{H}_{\mathrm{ss}}^{H}\right\}$, to design robust \textcolor{black}{TX} beamforming under the interference threshold. In Section \ref{Approximation}, we further utilize the information of terrestrial BS positions to approximate the integral-form interference channel term \( \boldsymbol{\Upsilon}^{\mathrm{int}}_{\mathrm{sg}} \), which avoids the complex integral calculations.
\section{Sum Rate Maximization Beamforming under Interference Threshold}\label{Algorithm-BC}

\vspace{-1mm}
\subsection{Problem Formulation}
\vspace{-1mm}

In this subsection, we design beamforming to maximize the weighted sum rate while ensuring that the average satellite-to-terrestrial UT interference remains below the threshold $I_{\rm{thr}}$, whose optimization problem can be formulated as follows

\begin{align}
	\begin{split}
	\max_{\mathbf{P}}&\quad  \sum_{k=1}^{K_\mathrm{S}} \mathbb{E}_{\mathbf{H}_{\mathrm{ss}}}\!\left\{a_k\log_{2}\!\left(\!1\! +\! \frac{\mathbf{p}_{k}^{H} \mathbf{h}_{\mathrm{ss}, k} \mathbf{h}_{\mathrm{ss}, k}^H \mathbf{p}_{k}}{\sum_{i \neq k} \mathbf{p}_{i}^{H} \mathbf{h}_{\mathrm{ss}, k} \mathbf{h}_{\mathrm{ss}, k}^H \mathbf{p}_{i}\! +\! \sigma_k^{2}}\!\right)\!\right\}, \\
		\text{ s.t. } & \quad \frac{1}{K_{\rm G}} \operatorname{Tr}\left\{\mathbf{P}^H\boldsymbol{\Upsilon}^{\mathrm {int}}_{\mathrm{sg}}\mathbf{P}\right\}  \leq I_{\mathrm{thr}}, \\
		& \quad \operatorname{Tr}\left\{\mathbf{P}\mathbf{P}^H\right\} \leq P_{\mathrm{T}},    \label{23}
	\end{split}
\end{align}
where $P_{\mathrm{T}}$ denotes the antenna power budget and $a_k$ represents the weight for the $k$-th satellite UT. 
\begin{ppn}\label{exp}
Problem \eqref{23} is equivalent to problem \eqref{23_1}, in the sense that the global optimal solution for the two problems is identical.
    \begin{align}
	\begin{split}
	\max_{\mathbf{P}}&\quad  \sum_{k=1}^{K_\mathrm{S}} \mathbb{E}_{\mathbf{B}}\!\left\{a_k\log_{2}\!\left(\!1\! +\! \frac{\mathbf{p}_{k}^{H} {\mathbf{b}}_{k} {\mathbf{b}}_{k}^H \mathbf{p}_{k}}{\sum_{i \neq k} \mathbf{p}_{i}^{H} {\mathbf{b}}_{k} {\mathbf{b}}_{k}^H \mathbf{p}_{i}\! +\! \sigma_k^{2}}\!\right)\!\right\}, \\
  {\rm{s.t.}}  & \quad \frac{1}{K_{\rm G}} \operatorname{Tr}\left\{\mathbf{P}^H\boldsymbol{\Upsilon}^{\mathrm {int}}_{\mathrm{sg}}\mathbf{P}\right\}  \leq I_{\mathrm{thr}}, \\
		& \quad \operatorname{Tr}\left\{\mathbf{P}\mathbf{P}^H\right\} \leq P_{\mathrm{T}},
        \label{23_1}
	\end{split}
\end{align}
where we define ${\mathbf{b}}_k = g_{k}^{\mathrm{ss}}\left(t,f\right)  \mathbf{v}_k^{\rm{ss}}  $  and \( {\mathbf{B}} = {[{\mathbf{b}}_{1}, \ldots, {\mathbf{b}}_{K_\mathrm{S}}]} \in \mathbb{C}^{M_\mathrm{S} \times K_\mathrm{S}} \) representing the satellite-to-satellite UT channel matrix which ignores the phase information.
\end{ppn}
\begin{pf}
	See Appendix \ref{app_exp}. 
 \end{pf}

\ppnref{exp} shows that we can obtain an optimal solution to problem \eqref{23} by solving problem \eqref{23_1}. However, problem \eqref{23_1} still constitutes a non-convex optimization problem, which poses significant challenges in obtaining the globally optimal solution.

\subsection{Alternating Optimization Algorithm based on MCQT}\label{Algorithm-B}
According to \textit{Jensen's inequality}, the ergodic sum rate, i.e. the objective function in \eqref{23_1}, is lower bounded by \cite{10107609}
\begin{align}
	\begin{split}
  & \sum_{k=1}^{K_\mathrm{S}} \mathbb{E}_{\mathbf{B}}\left\{a_k\log_{2}\left(1 +\frac{\mathbf{p}_{k}^{H} {\mathbf{b}}_{k} {\mathbf{b}}_{k}^H \mathbf{p}_{k}}{\sum_{i \neq k} \mathbf{p}_{i}^{H} {\mathbf{b}}_{k} {\mathbf{b}}_{k}^H \mathbf{p}_{i} +\sigma_k^{2}}\right)\right\} \\& \geq  \sum_{k=1}^{K_{\rm S}}a_k\log_2 \left(\!1+\frac{ \mathbf{p}_{k}^{H}\bar{\mathbf{h}}_k\bar{\mathbf{h}}_k^H\mathbf{p}_{k}}{\sum_{i=1}^{K_{\rm S}} \mathbf{p}_{i}^{H}{\boldsymbol{\Upsilon}}_{\mathrm{ss},k} \mathbf{p}_{i}-\mathbf{p}_{k}^{H}\bar{\mathbf{h}}_k\bar{\mathbf{h}}_k^{H} \mathbf{p}_{k}+\sigma_k^2}\right),
   \label{23_1_1}
	\end{split}	
\end{align}
 where we define $\bar{\mathbf{h}}_k=\mathbb{E}\left\{{\mathbf{b}}_{k} \right\}={\bar\gamma_{k}^{\mathrm{ss}}}\mathbf{v}_{k}^{\mathrm{ss}}$ and ${\boldsymbol{\Upsilon}}_{\mathrm{ss},k}=\mathbb{E}\left\{{\mathbf{b}}_{k}{\mathbf{b}}_{k}^H\right\}=(\gamma_{k}^{\mathrm{ss}})^2\mathbf{v}_{k}^{\mathrm{ss}}{\mathbf{v}_{k}^{\mathrm{ss}}}^H$.
As such, we reformulate the direct maximization of the ergodic sum rate as the maximization of its lower bound. Since this problem is a standard concave-convex multiple-ratio fractional programming problem, we apply the multidimensional complex quadratic transformation (MCQT) to tackle it. This transformation decouples the numerator and denominator of each SINR through MCQT decoupling \cite{8314727}. Consequently, the original WSR problem \eqref{23} is transformed into
\begin{align}
	\begin{split}
		\max _{\boldsymbol{\xi},\mathbf{P}} & \quad f(\boldsymbol{\xi},\mathbf{P}),\\
		\text{ s.t. } & \quad \frac{1}{K_{\rm G}} \operatorname{Tr}\left\{\mathbf{P}^H\boldsymbol{\Upsilon}^{\mathrm {int}}_{\mathrm{sg}}\mathbf{P}\right\}\leq I_{\mathrm{thr}}, \\
		&\quad  \operatorname{Tr}\left\{\mathbf{P}\mathbf{P}^H\right\} \leq P_{\mathrm{T}},
		\label{26}
	\end{split}	
\end{align}
where ${\boldsymbol \xi \in \mathbb{C}^{K_\mathrm{S} \times 1}}$ denotes the auxiliary variable, and the objective function given by \eqref{24_1_1} is convex with respect to $\boldsymbol{\xi}$ and $\mathbf{P}$, respectively. 

\begin{figure*}[htbp]
	\normalsize
\begin{align}
	\begin{split}
         f(\boldsymbol{\xi}, \mathbf{P})=  \sum_{k=1}^{K_\mathrm{S}} a_k\log_2 \left(1+2 \operatorname{Re}\left\{{\xi}_{k}^{*} \bar{\mathbf{h}}_k^H\mathbf{p}_{k}\right\} -{\xi}_{k}^{*}\left(\sum_{i=1}^{K_{\rm S}} \mathbf{p}_{i}^{H}{\boldsymbol{\Upsilon}}_{\mathrm{ss},k} \mathbf{p}_{i}-\mathbf{p}_{k}^{H}\bar{\mathbf{h}}_k\bar{\mathbf{h}}_k^{H} \mathbf{p}_{k}+\sigma_k^2 \right) {\xi}_{k}\right).
		\label{24_1_1}
	\end{split}	
\end{align}
\hrulefill
\vspace{-3mm}
\end{figure*}
\begin{algorithm}[t]
	\caption{WSR-MCQT-Based \textcolor{black}{TX} Beamforming}
	\label{A1}
		\begin{algorithmic}[1]
			\STATE \textbf{Input:}\!\!  $\left\{\mathbf{v}_k^{\rm{ss}}, \gamma_k^{\rm{ss}},\bar\gamma_k^{\rm{ss}}, \sigma_k^2\right\}_{k=1}^{K_{\rm S}}$, $\boldsymbol{\Upsilon}^{\mathrm{int}}_{\mathrm{sg}}$, $I_{\mathrm{thr}}$, $P_{\mathrm{T}}$, ${\rm Iter}_{\rm max}$, $\delta$.
			\STATE Initialize the beamforming matrix \(\mathbf{P}\), ${f}^{(0)}$ and $n=1$.
 		\STATE \textbf{while} $n<{\rm Iter}_{\rm max}$ \textbf{do}
			\STATE \quad Calculate auxiliary variable $\left\{{\xi}_{k}^{\star}\right\}_{k=1}^{K_{\rm S}}$ from \eqref{25}. 
			\STATE \quad
			Obtain $\mathbf{P}^{\star}$ by solving problem \eqref{24_2}.
			\STATE \quad Calculate ${f}^{(n)}$ and $I_{\mathrm{avg}}^{(n)}$.
			\STATE \quad \textbf{if} $|	{f}^{(n)} -	{f}^{(n-1)} |<\delta $ \textbf{and} $ I_{\mathrm{avg}}^{(n)} < I_{\mathrm{thr}}$ 
			\STATE \quad \quad \textbf{break}
			\STATE \quad \textbf{end if}
			\STATE \quad $n = n + 1$.
			\STATE \textbf{end while}
			\STATE \textbf{Output:} ${\boldsymbol \xi}^{\star}$ and $\mathbf{P}^{\star}$.
		\end{algorithmic}
\end{algorithm}

On the basis of this transformation, we propose an alternating optimization algorithm to address this problem. Specifically, when matrix \( \mathbf{P} \) is fixed, the problem becomes an unconstrained convex optimization as $\mathop{\rm max}\limits_{\boldsymbol \xi} \, {f}(\boldsymbol \xi,\mathbf{P})$.
Here, the optimal solution ${\xi} _{k}^{\star}$ can be obtained by setting the  \( \nabla_{{\boldsymbol \xi}^*} {f}(\boldsymbol \xi) \) to zero, as shown below
\begin{align}
{\xi}_{k}^{\star}&=\frac{{\bar{\mathbf{h}}_k^{H}\mathbf{p}_{k}}}{\sum_{i=1}^{K_{\rm S}} \mathbf{p}_{i}^{H}{\boldsymbol{\Upsilon}}_{\mathrm{ss},k} \mathbf{p}_{i}-\mathbf{p}_{k}^{H}\bar{\mathbf{h}}_k\bar{\mathbf{h}}_k^{H} \mathbf{p}_{k}+\sigma_k^2}
\\&=\frac{{({\bar\gamma_{k}^{\rm{ss}}})^*}(\mathbf{v}_{k}^{\rm{ss}})^H\mathbf{p}_{k}}{\sum_{i=1}^{K_{\rm S}} (\gamma_{k}^{\rm{ss}})^2\mathbf{p}_{i}^{H}\mathbf{v}_{k}^{\rm{ss}}(\mathbf{v}_{k}^{\rm{ss}})^H \mathbf{p}_{i}-|{\bar\gamma_{k}^{\rm{ss}}}|^2\mathbf{p}_{k}^{H}\mathbf{v}_{k}^{\rm{ss}}(\mathbf{v}_{k}^{\rm{ss}})^H \mathbf{p}_{k}+\sigma_k^2}.
\label{25}
\end{align}

With the auxiliary variable \(\boldsymbol{\xi}^{\star}\) fixed, the original optimization problem in \eqref{26} can be transformed into
\begin{align}
	\begin{split}
		\max _{\mathbf{P}} & \quad {f} (\mathbf{P}),\\
		\text{ s.t. } & \quad \frac{1}{K_{\rm G}} \operatorname{Tr}\left\{\mathbf{P}^H\boldsymbol{\Upsilon}^{\mathrm {int}}_{\mathrm{sg}}\mathbf{P}\right\}  \leq I_{\mathrm{thr}}, \\
		&\quad  \operatorname{Tr}\left\{\mathbf{P}\mathbf{P}^H\right\} \leq P_{\mathrm{T}}.
		\label{24_2}
	\end{split}	
\end{align}
The above problem is a convex problem and can be solved with generic convex optimization algorithms, e.g., the interior point algorithm \cite{1570864}. The optimization of \( \boldsymbol \xi \) and \( \mathbf{P} \) alternates until convergence, and this process is summarized in \algref{A1}.

The computational complexity of \algref{A1} is dominated by Step 5 and Step 6. Specifically, Step 5 solving for \(\mathbf{P}^{\star}\) using the interior-point method performs with a complexity order of \(\mathcal{O}\left(T_A \left(M_{\rm S}^3 K_{\rm S}^{3}\right)\right)\), assuming a maximum of \(T_A\) inner loop iterations; Step 6 performs the matrix multiplication for the calculation of the objective function and the computation of average interference, involving both integral calculation and matrix multiplication, has a complexity order of \(\mathcal{O}\left(M_{\rm S}K_{\rm S}^2 +M_{\mathrm{S}}^{2} N_{\rm G} N_r N_{\phi} \right)\). Herein, \( N_r \) and \( N_{\phi} \) denote the numbers of discrete samples for the integral over the terrestrial BS radius \( r \) and the polar angle \( \phi \), respectively.
 Thus, the total complexity order of \algref{A1} is \(\mathcal{O}\left(T_A \left(M_{\rm S}^3 K_{\rm S}^{3}\right)+M_{\mathrm{S}}^{2}N_{\rm G} N_r N_{\phi}\right)\) per iteration.

 \vspace{-1mm}
\subsection{WMMSE Algorithm under Interference Threshold}\label{Algorithm-C}
\vspace{-1mm}

Although the scheme in \secref{Algorithm-B} achieves outstanding performance, the subproblem in \eqref{24_2} still incurs high computational complexity due to the large dimension of \(\mathbf{P}\). Therefore, in this subsection, we propose an iterative algorithm derived from the equivalence between the WSR and WMMSE problems, termed WSR-WMMSE equivalence (WWE), which leverages the KKT conditions of the WMMSE problem to facilitate a solving process with lower computational complexity. With the definitions $\mathbf{u} \in \mathbb{C}^{K_{\mathrm{S}} \times 1}$ and $\mathbf{w} \in \mathbb{C}^{K_{\mathrm{S}} \times 1}$, the following WMMSE problem
\begin{align}
	\begin{split}
		\underset{\mathbf{u}, \mathbf{w}, \mathbf{P}}{\min} &\quad \sum_{k=1}^{K_\mathrm{S}} \left({w}_{k} {e}_{k}-\log  {w}_{k}\right) , \\
		\text {s.t.}& \quad \frac{1}{K_{\rm G}} \operatorname{Tr}\left\{\mathbf{P}^H\boldsymbol{\Upsilon}^{\mathrm {int}}_{\mathrm{sg}}\mathbf{P}\right\} \leq I_{\mathrm{thr}}, \\
		& \quad \operatorname{Tr}\left\{\mathbf{P}\mathbf{P}^H\right\} \leq P_{\mathrm{T}},
		\label{27}
	\end{split}	
\end{align}
is equivalent to the WSR problem in \eqref{23} with the lower bound objective in \eqref{23_1_1}, in the sense that the global optimal solution $\mathbf{P}$ for the two problems are identical  \cite{10440321,5756489}. The MSE $e_k$ can be expressed as 
\begin{align}
	\begin{split}
&{e}_{k} = \mathbb{E}_{{\mathbf{b}}_k, \mathbf{x}_{\mathrm{s}}, {n}_k}\left\{\left|{u}_{k}\left({\mathbf{b}}_k^H\mathbf{P}\mathbf{x}_{\mathrm{s}}+{n}_k\right)-{x}_{k}\right|^{2}\right\}\\
        &= \!\left|u_k \right|^{2}\!\!\left(\sum_{i = 1}^{K_{\rm S}} \mathbf{p}_{i}^{H} {\boldsymbol{\Upsilon}}_{\mathrm{ss},k}\mathbf{p}_{i}+\sigma_k^2\!\right)\! -\!u_k^*\mathbf{p}_k^H\bar{\mathbf{h}}_{k}- u_k\bar{\mathbf{h}}_{k}^H\mathbf{p}_k+1.
		\label{29}
	\end{split}	
\end{align}
Therefore, with \(\mathbf{P}\) and \(\mathbf{w}\) fixed, the optimal value of \(u_k^{\star}\) can be determined by setting the derivative of \(e_k(u_k)\) with respect to \(u_k^*\) equal to zero
\begin{align}
	\begin{split}
		{u}_{k}^{\star}  &=\frac{\mathbf{p}_k^H\bar{\mathbf{h}}_{k}}{\sum_{i = 1}^{K_{\rm S}} \mathbf{p}_{i}^{H} {\boldsymbol{\Upsilon}}_{\mathrm{ss},k} \mathbf{p}_{i} +\sigma_k^{2}}\\
&=\frac{{\bar\gamma_{k}^{\rm ss}}\mathbf{p}_{k}^H\mathbf{{v}}_{k}^{\rm ss}}{\sum_{i =1}^{K_{\rm S}} (\gamma_{k}^{\rm ss})^2\mathbf{p}_{i}^{H} \mathbf{v}_{k}^{\rm ss}{\mathbf{v}_{k}^{\rm ss}}^H \mathbf{p}_{i} +\sigma_k^{2}}.
		\label{28}
	\end{split}	
\end{align}
 At this point, the value of \( e_k \) can be computed as
\begin{align}
	\begin{split}
e_k
&=1-\frac{\mathbf{p}_{k}^H \bar{\mathbf{h}}_{k}\bar{\mathbf{h}}_{k}^H \mathbf{p}_{k}}{\sum_{i = 1}^{K_{\rm S}} \mathbf{p}_{i}^{H} {\boldsymbol{\Upsilon}}_{\mathrm{ss},k} \mathbf{p}_{i} +\sigma_k^{2}}\\
&=1-\frac{|{\bar\gamma_{k}^{\rm ss}}|^2 \mathbf{p}_{k}^H\mathbf{v}_{k}^{\rm ss}{\mathbf{v}_{k}^{\rm ss}}^H\mathbf{p}_{k}}{\sum_{i =1}^{K_{\rm S}} (\gamma_{k}^{\rm ss})^2\mathbf{p}_{i}^{H} \mathbf{v}_{k}^{\rm ss}{\mathbf{v}_{k}^{\rm ss}}^H \mathbf{p}_{i} +\sigma_k^{2}}.
\label{}
\end{split}	
\end{align}

The equivalence between the WSR and WMMSE problems can be established when the KKT conditions of both problems are simultaneously satisfied \cite{4712693}. By equating the gradients of their respective Lagrangian functions, the optimal MSE-weights \(\{w_k\}_{k=1}^{K_{\rm S}}\) can be derived for given \(\mathbf{P}\) and \(\mathbf{u}^{\star}\) as \cite{5756489}
\begin{align}
	\begin{split}
		{w}_{k}^{\star}=a_k{e}_{k}^{-1}.
		\label{31}
	\end{split}	
\end{align}

Then we fix vectors $\mathbf{w}^{\star}$ and $\mathbf{u}^{\star}$, the optimization problem is reduced to
\begin{align}
	\begin{split}
		\underset{\mathbf{P}}{\min} &\quad \epsilon(\mathbf{P})=\sum_{k=1}^{K_\mathrm{S}} {w}_{k} {e}_{k}(\mathbf{P}) , \\
		\text { s.t. }& \quad \frac{1}{K_{\rm G}} \operatorname{Tr}\left\{\mathbf{P}^H\boldsymbol{\Upsilon}^{\mathrm {int}}_{\mathrm{sg}}\mathbf{P}\right\} \leq I_{\mathrm{thr}}, \\
		& \quad \operatorname{Tr}\left\{\mathbf{P}\mathbf{P}^H\right\} \leq P_{\mathrm{T}},
		\label{WSR-WMMSE}
	\end{split}	
\end{align}
whose Lagrange function can be derived as
\begin{align}
	\begin{split}
\mathcal{L}\left(\mathbf{P}\right)=&\sum_{k=1}^{K_{\mathrm{S}}}{w}_{k} {e}_{k}(\mathbf{P}) +\lambda \left( \operatorname{Tr}\left\{\mathbf{P}\mathbf{P}^H\right\} - P_{\mathrm{T}}\right)\\
	&+\mu\left(\frac{1}{K_{\rm G}} \operatorname{Tr}\left\{\mathbf{P}^H\boldsymbol{\Upsilon}^{\mathrm {int}}_{\mathrm{sg}}\mathbf{P}\right\} - I_{\mathrm{thr}}\right),
		\label{32}
\end{split}	
\end{align}
where $\lambda$ and $\mu$ are the Lagrange multipliers. By setting \( \nabla_{\mathbf{P}^*} \mathcal{L} = \mathbf{0} \), we formulate the closed-form iterative expression for \( \mathbf{P}^{\star} \) as
\begin{align}
	\textstyle\mathbf{P}^{\star}(\lambda,\mu) =\left(\hat{\boldsymbol{\Upsilon}}_{\mathrm{ss}}\!+\!\lambda \mathbf{I}+\frac{\mu}{K_{\rm G}}  \boldsymbol{\Upsilon}^{\mathrm {int}}_{\mathrm{sg}}\right)^{-1}\mathbf{W}\mathbf{U}^H\bar{\mathbf{H}}_{\rm{ss}} ,
    \label{33}
\end{align}
where we denote \(\hat{\boldsymbol{\Upsilon}}_{\mathrm{ss}} = \sum_{k=1}^{K_{\rm S}} u_k^*w_ku_k\mathbb{E}\left\{{\mathbf{b}}_k{\mathbf{b}}_k^H\right\} = \sum_{k=1}^{K_{\mathrm{S}}} u_k^*w_ku_k (\gamma_{k}^{\rm ss})^2 \mathbf{v}_{k}^{\rm ss} {\mathbf{v}_{k}^{\rm ss}}^H\). Moreover, we define matrices \( \mathbf{W}^{\star}=\mathrm{diag}\left\{[w_1,, \ldots,w_{K_{\rm S}}]\right\} \), \( \mathbf{U}^{\star}=\mathrm{diag}\left\{[u_1,, \ldots,u_{K_{\rm S}}]\right\} \) and $\bar{\mathbf{H}}_{\mathrm{ss}}=\mathbb{E}\left\{\mathbf{B}\right\}$. 
\begin{algorithm}[t]
	\caption{WWE-Based \textcolor{black}{TX} Beamforming}
	\label{A2}
		\begin{algorithmic}[1]
			\STATE \textbf{Input:} \!\!$\left\{\mathbf{v}_k^{\rm{ss}}, \gamma_k^{\rm{ss}},\bar\gamma_k^{\rm{ss}},\sigma_k^2\right\}_{k=1}^{K_{\rm S}}$, $\boldsymbol{\Upsilon}^{\mathrm{int}}_{\mathrm{sg}}$, $I_{\mathrm{thr}}$, $P_{\mathrm{T}}$, ${\rm Iter}_{\rm max}$, $\delta$.
			\STATE Initialize the beamforming matrix \(\mathbf{P}\), $\epsilon^{(0)}$ and $n=1$.
			\STATE \textbf{while} $n<{\rm Iter}_{\rm max}$ \textbf{do}
			\STATE \quad Calculate  $\left\{{u}_{k}^{\star}\right\}_{k=1}^{K_{\rm S}}$  and $\left\{{w}_{k}^{\star}\right\}_{k=1}^{K_{\rm S}}$ from \eqref{28} and \eqref{31}. 
			\STATE \quad Solve $\lambda^{(n)}$ and $\mu^{(n)}$ from problem \eqref{lambda-mu}.
		    \STATE \quad Calculate  $\mathbf{P}^{\star}$ from \eqref{33}. 
        	\STATE \quad Calculate $\epsilon^{(n)}$ and $I_{\mathrm{avg}}^{(n)}$.
            \STATE \quad \textbf{if} $|	\epsilon^{(n)} -\epsilon^{(n-1)} |<\delta $ \textbf{and} $ I_{\mathrm{avg}}^{(n)} < I_{\mathrm{thr}}$ 
			\STATE \quad \quad \textbf{break}
			\STATE \quad \textbf{end if}
			\STATE \quad $n = n + 1$.
			\STATE \textbf{end while}
			\STATE \textbf{Output:} $\mathbf{P}^{\star}$.
		\end{algorithmic}
\end{algorithm}
By substituting this expression into problem \eqref{WSR-WMMSE}, it can be reduced as the following optimization problem in terms of \(\lambda\) and \(\mu\)
\begin{align}
	\begin{split}
		\underset{\lambda, \mu}{\min} &\quad \epsilon(\lambda,\mu)=\sum_{k=1}^{K_\mathrm{S}} {w}_{k} {e}_{k}(\lambda, \mu) , \\
		\text { s.t. }& \quad \frac{1}{K_{\rm G}} \operatorname{Tr}\left\{\mathbf{P}^H(\lambda, \mu)\boldsymbol{\Upsilon}^{\mathrm {int}}_{\mathrm{sg}}\mathbf{P}(\lambda, \mu)\right\} \leq I_{\mathrm{thr}}, \\
		& \quad \operatorname{Tr}\left\{\mathbf{P}(\lambda, \mu)\mathbf{P}^H(\lambda, \mu)\right\} \leq P_{\mathrm{T}}.
		\label{lambda-mu}
	\end{split}	
\end{align}

 Although the above problem is non-convex, its low-dimensional optimization variable allows us to solve it efficiently by conventional iterative algorithms. After this, by bringing $\lambda^{\star}$ and $\mu^{\star}$ back to formula \eqref{33}, we can get the result of $\mathbf{P}^{\star}$ without loss of performance.
 It is worth noting that, within the alternating optimization framework, rather than searching for the optimal solution of $\mathbf{P}$ in \eqref{WSR-WMMSE}, we can focus on finding feasible solutions that lead to a decreasing objective function. This approach provides further reductions in the computational complexity, while also ensuring gradual improvement of the original optimization objective.
 In conclusion, we optimize the Lagrange multipliers \(\lambda\) and \(\mu\) in each iteration, and update \(\mathbf{u}^{\star}\), \(\mathbf{w}^{\star}\) and \(\mathbf{P}^{\star}\) alternately until convergence, summarized in \algref{A2}.

The computational complexity of \algref{A2} primarily arises from Steps 5 and 6. Specifically, Step 5 solving for \(\lambda\) and \(\mu\) using the interior point method involves a complexity order of \(\mathcal{O}(T_B )\) where \(T_B\) represents the maximum number of inner iterations; Step 6, which involves calculating \(\mathbf{P}^{\star}\), has a complexity order of \(\mathcal{O}(M_{\rm S}^2 K_{\rm S} + M_{\rm S}^2 N_{\rm G} N_r N_{\phi} + M_{\rm S}^3)\). Consequently, the overall complexity order of \algref{A2} per iteration is {\(\mathcal{O}(T_B + M_{\rm S}^2K_{\rm S} + M_{\mathrm{S}}^{2} N_{\rm G} N_r N_{\phi}+M_{\rm S}^3 )\)}, which represents a significant reduction compared to \algref{A1}.

\section{Closed-Form \textcolor{black}{TX} Beamforming Based on MMSE for Interference Avoidance}\label{Algorithm-A}
As another linear beamforming design criterion, MMSE is widely adopted due to its significant enhancement of demodulation performance and the availability of closed-form optimal solutions in many cases. In this section, we revisit the MMSE beamforming design under the constraint of inter-system interference and based on the available sCSI.

\vspace{-1mm}
\subsection{Problem Formulation}
\vspace{-1mm}
The \textcolor{black}{TX} beamforming optimization problem with the MMSE criterion under the interference threshold is formulated as follows
\begin{align}
	\begin{split}
		\min _{\mathbf{P},\beta} &  \quad \mathbb{E}_{\mathbf{H}_\mathrm{ss},\mathbf{x}_\mathrm s, \mathbf{n}_\mathrm s}\left\{\left\|\frac{\boldsymbol \Psi_{\rm s}\left(\mathbf{H}_\mathrm{ss}^H\mathbf{P}\mathbf{x}_\mathrm s+\mathbf{n}_\mathrm s\right)}{\beta} -\mathbf{x}_\mathrm s\right\|_{2}^{2}\right\}, \\
		\text { s.t. }& \quad \frac{1}{K_{\rm G}} \operatorname{Tr}\left\{\mathbf{P}^H\boldsymbol{\Upsilon}^{\mathrm {int}}_{\mathrm{sg}}\mathbf{P}\right\} \leq I_{\mathrm{thr}}, \\
		& \quad \operatorname{Tr}\left\{\mathbf{P}\mathbf{P}^H\right\} \leq P_{\mathrm{T}}.
	\end{split}
	\label{}
\end{align}
Herein, $\boldsymbol \Psi_{\rm s}=\operatorname{diag}\left\{[\mathrm{e}^{j\psi_1}, \ldots, \mathrm{e}^{j\psi_{K_{\rm s}}}]\right\}$ represents the phase compensation for delay and Doppler shifts at the receiver with \(\psi_k=2 \pi\left(t v_k^{\rm{sat}}- f \tau_{k}^{\rm{min}}\right)\). Compared to the difficulty of phase compensation at the transmitter, the receiver can accurately estimate phase errors based on downlink pilot signals to assist with compensation  \cite{10066300,hou2024}.
To simplify the problem, we employ the penalty function method to mitigate interference \cite{Wang2013PenaltyFP}. Specifically, we introduce a penalty term \(\bar\varsigma  \operatorname{Tr}\left\{\mathbf{P}^H\boldsymbol{\Upsilon}_{\mathrm{sg}}^{\rm int}\mathbf{P}\right\}\) into the objective function, transforming the problem into a convex optimization problem subject to power budget. Here, \(\bar \varsigma \) represents the penalty factor, and for the sake of derivation, we set \(\bar \varsigma  = \frac{\varsigma }{\beta^2}\), allowing us to rewrite the optimization problem as follows
\begin{align}
	\begin{split}
		\min _{\mathbf{P},\beta} & \, \mathbb{E}_{\mathbf{H}_\mathrm{ss},\mathbf{x}_\mathrm s, \mathbf{n}_\mathrm s}\!\!\left\{\!\left\|\frac{\!\boldsymbol \Psi_{\rm s}\!\left(\mathbf{H}_\mathrm{ss}^H\mathbf{P}\mathbf{x}_\mathrm s\!+\!\mathbf{n}_\mathrm s\right)}{\beta}\!-\!\mathbf{x}_\mathrm s\right\|_{2}^{2}\!\right\}\!
		\!+\!\bar\varsigma  \operatorname{Tr}\!\left\{\!\mathbf{P}^H\!\boldsymbol{\Upsilon}^{\mathrm {int}}_{\mathrm{sg}}\mathbf{P}\!\right\}\!, \\
		\text {s.t.} & \,\operatorname{Tr}\left\{\mathbf{P}\mathbf{P}^H\right\}\leq P_{\mathrm{T}}.
	\end{split}
	\label{12}
\end{align}
\begin{ppn}\label{closed-form}
The following solution achieves the optimum of problem \eqref{12}, where
\begin{align}
&\mathbf{P}^{\star}=\beta^{\star}\left(\boldsymbol{\Upsilon}_{\mathrm{ss}}+\varsigma \boldsymbol{\Upsilon}^{\mathrm {int}}_{\mathrm{sg}}+ \frac{K_{\mathrm{S}}\sigma_{\rm s}^2}{P_\mathrm T}\mathbf{I}\right)^{-1} \bar{\mathbf{H}}_{\mathrm{ss}},\label{20}\\
&\ \ \textstyle\beta^{\star} = \sqrt{\frac{P_\mathrm{T}}{\left\|{\left(\boldsymbol{\Upsilon}_{\mathrm{ss}}+\varsigma \boldsymbol{\Upsilon}^{\mathrm {int}}_{\mathrm{sg}}+ \frac{K_{\mathrm{S}}\sigma_{\rm s}^2}{P_\mathrm T}\mathbf{I}\right)^{-1}\bar{\mathbf{H}}_\mathrm{s s}}\right\|_F^2}},
\end{align}  
where \({\boldsymbol{\Upsilon}}_{\mathrm{ss}} \triangleq\mathbb{E}\left\{\mathbf{H}_{\mathrm{ss}}\mathbf{H}_{\mathrm{ss}}^H\right\}=  \sum_{k=1}^{K_{\mathrm{S}}} (\gamma_{k}^{\rm ss})^2\mathbf{v}_{k}^{\rm ss} {\mathbf{v}_{k}^{\rm ss}}^H\) and it is assumed that $\sigma_k^2=\sigma_{\rm s}^2$ for all UTs. 

\end{ppn}
\begin{pf}
	See Appendix \ref{app_closed-form}. 
\end{pf}
\begin{remark}
	When both \(\mathbf{W}\) and \(\mathbf{U}\) are identity matrices, and by forcing \(\lambda =\frac{K_{\mathrm{S}}\sigma_{\rm s}^2}{P_\mathrm T}\) and \(\mu =\varsigma  K_{\rm G} \), the beamforming matrix in formula \eqref{33} can degenerate into the closed-form MMSE solution in formula \eqref{20}.
\end{remark}
\subsection{Closed-Form Beamforming under Interference Threshold}
The selection of \( \varsigma  \) is crucial, as it effectively represents the trade-off between MMSE performance and interference management capability. Specifically, the relationship is summarized in \ppnref{dI-dk}.
\begin{ppn}\label{dI-dk}
	For the closed-form solution of \(\mathbf{P}\) in \eqref{20}, the derivative of the total interference \(I_\mathrm{sg}\) with respect to \(\varsigma \) is non-positive. 
	\begin{align}
		\begin{split} 
			\nabla_{\varsigma } I_\mathrm{sg}(\varsigma )=\frac{\partial I_\mathrm{sg}(\varsigma )}{\partial \varsigma }=\frac{\partial\operatorname{Tr}\left\{\mathbf{P}^H(\varsigma )\boldsymbol{\Upsilon}^{\mathrm {int}}_{\mathrm{sg}}\mathbf{P}(\varsigma )\right\}}{\partial \varsigma } \leq 0.
		\end{split}
		\label{22}
	\end{align}
\end{ppn}
\begin{pf}
	See Appendix \ref{app_dI-dk}. 
\end{pf}

Given that a higher value of \( \varsigma  \) results in a lower interference power \( I_\mathrm{sg} \), the value of \( \varsigma  \) corresponding to a fixed average interference threshold \( I_\mathrm{thr} \) and a specific signal-to-noise ratio (SNR) can be determined through a bisection method. The detailed algorithm is described in \algref{A3}.

\begin{algorithm}[t]
	\caption{Closed-Form \textcolor{black}{TX} Beamforming}
	\label{A3}
		\begin{algorithmic}[1]
			\STATE \textbf{Input:} $\left\{\mathbf{v}_k^{\rm{ss}},\gamma_k^{\rm{ss}},\bar\gamma_k^{\rm{ss}}\right\}_{k=1}^{K_{\rm S}}$, $\boldsymbol{\Upsilon}^{\mathrm{int}}_{\mathrm{sg}}$, $I_{\mathrm{thr}}$, $P_{\mathrm{T}}$, $\sigma_{\rm s}^2$.
			\STATE Find \(\varsigma \) for given \(I_\mathrm{thr}\) and $\rm{SNR}$ via a bisection method.
\STATE $\beta^{\star} =
\sqrt{ {P_\mathrm{T}}/{
\|( \boldsymbol{\Upsilon}_{\mathrm{ss}} + \varsigma \boldsymbol{\Upsilon}^{\mathrm{int}}_{\mathrm{sg}} + \frac{K_{\mathrm{S}} \sigma_{\rm s}^2}{P_\mathrm{T}} \mathbf{I} )^{-1}
\bar{\mathbf{H}}_{\mathrm{ss}} \|_F^2 } }$.
\STATE  $\mathbf{P}^{\star}=\beta^{\star}\left(\boldsymbol{\Upsilon}_{\mathrm{ss}}+\varsigma \boldsymbol{\Upsilon}^{\mathrm {int}}_{\mathrm{sg}}+ \frac{K_{\mathrm{S}}\sigma_{\rm s}^2}{P_\mathrm T}\mathbf{I}\right)^{-1} \bar{\mathbf{H}}_\mathrm{s s}$.
\STATE \textbf{Output:} $\beta^\star$ and $\mathbf{P}^{\star}$.
\end{algorithmic}
\end{algorithm}

Although the bisection method requires iterative calculations of the beamforming matrix when performing \textcolor{black}{TX} beamforming over continuous time slots or beamforming periods, we can significantly reduce the frequency of executing the bisection method by predicting the variation of \(\varsigma \). This approach enables the beamforming scheme to be approximated as a closed-form computation.
Upon analyzing the solution of \(\mathbf{P}^{\star}\), the overall complexity order is \(\mathcal{O}\left(M_{\mathrm{S}}^{2} N_{\rm G} N_r N_{\phi} + M_{\mathrm{S}}^{3}\right)\). Compared to the iterative solutions for the beamforming matrix discussed in the previous section, this closed-form beamforming solution significantly reduces the computational complexity.

\section{Position-Aided Interference Approximation in \textcolor{black}{TX} Beamforming Design}\label{Approximation}

In the proposed schemes, the \textcolor{black}{TX} beamforming relies on the integral term \(\boldsymbol{\Upsilon}_{\mathrm{sg}}^\mathrm{int}\). However, this integral term is computationally intensive and requires a real-time distribution of terrestrial users, which poses practical challenges. To address these issues, we leverage the information of terrestrial BS positions to approximate \(\boldsymbol{\Upsilon}_{\mathrm{sg}}^\mathrm{int}\), thereby eliminating reliance on the integral of terrestrial user distribution and significantly reducing the overall computational complexity of the beamforming schemes. We call this approximation process the terrestrial BS position-aided (PA) method. \begin{figure}[tp]
	\centering
\includegraphics[width=0.97\linewidth,trim=0.5cm 0.7cm 0.5cm 0.7cm,clip]{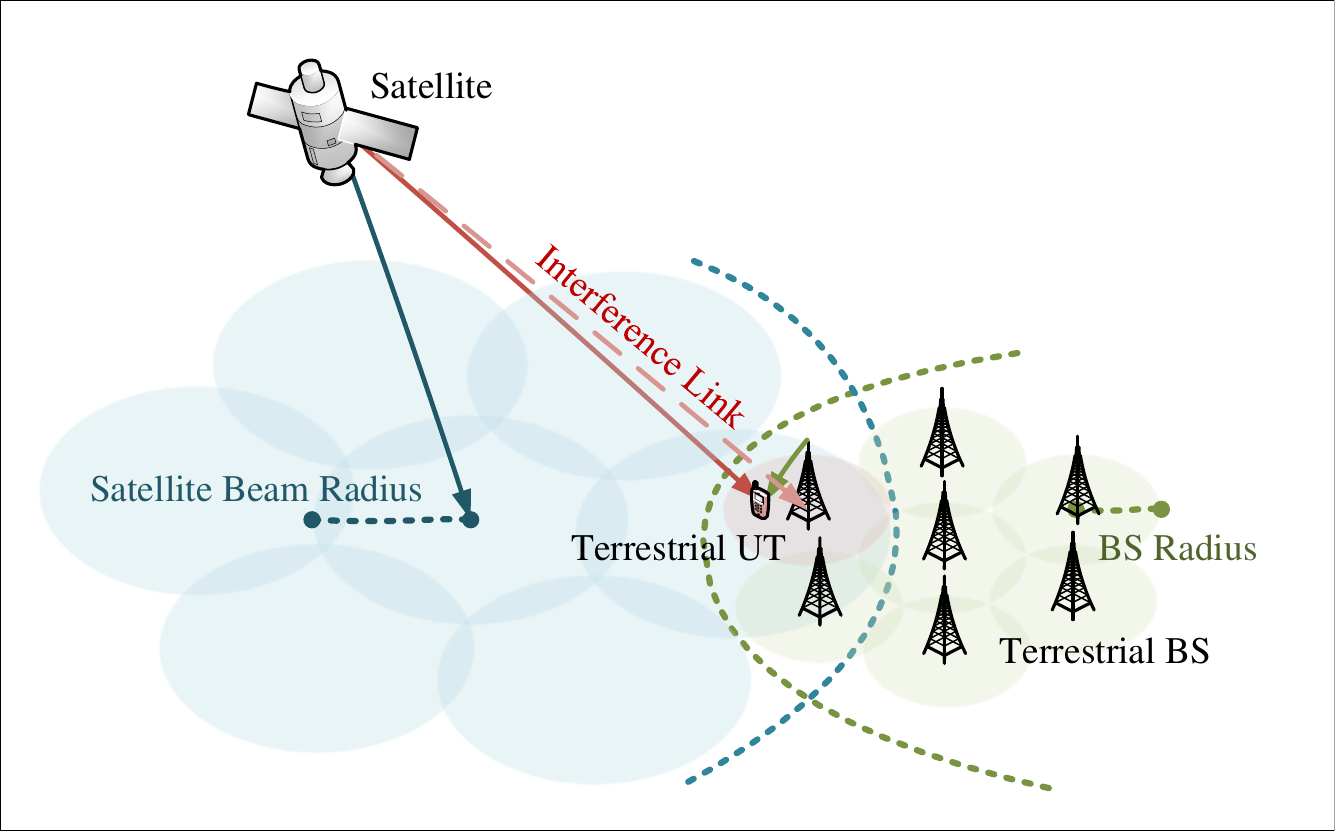}
\caption{Approximation of terrestrial UTs with terrestrial BS position.}
	\label{bs_aided}
\end{figure}\textcolor{black}{Specifically, positions of terrestrial BSs are assumed to be prior knowledge due to their fixed and planned deployment, allowing the satellite to compute the angular parameters (e.g., azimuth and elevation) via geometric relationships. Based on this, the satellite-to-BS channel is estimated with the Rician factor inferred from angular information\cite{3GPPTR38.811}.}

As depicted in \figref{bs_aided}, the coverage area of the terrestrial BS cell is significantly smaller than the size of satellite beams \cite{10179219}. Therefore, we approximate the satellite-to-terrestrial UT interference channel as the corresponding satellite-to-terrestrial BS channel. We define \( \tilde{\mathbf{H}}_{\mathrm{sg}} \) to represent the channel from the satellite to the terrestrial BSs and the approximation of \(\boldsymbol{\Upsilon}_{\mathrm{sg}}^\mathrm{int}\) can be expressed as \(\tilde{\boldsymbol{\Upsilon}}_{\mathrm{sg}}=\bar{K}_\mathrm{G} \mathbb{E}\left\{\tilde{\mathbf{H}}_{\mathrm{sg}} \tilde{\mathbf{H}}_{\mathrm{sg}}^H\right\}\) where \(\bar{K}_\mathrm{G}\) is the number of UTs per terrestrial BS. 
The original constraint of average interference power can be approximated as
\begin{align}
	\begin{split}
& \frac{1}{K_{\rm G}} \operatorname{Tr}\left\{\mathbf{P}^H\bar{K}_\mathrm{G} \mathbb{E}\left\{\tilde{\mathbf{H}}_{\mathrm{sg}}\tilde{\mathbf{H}}_{\mathrm{sg}}^H \right\}\mathbf{P}\right\}\\
& =\frac{1}{K_{\rm G}} \operatorname{Tr}\left\{\mathbf{P}^H \tilde{\boldsymbol{\Upsilon}}_{\mathrm{sg}}\mathbf{P}\right\}  \leq I_{\mathrm{thr}}.
	\end{split}
	\label{approx_bs}
\end{align}

 The other steps in \secref{Algorithm-BC} remain unchanged, except for the formula in \eqref{33} based on the equivalence between the WSR and WMMSE problems, which needs to be modified as
\begin{align}
\mathbf{P}_{\rm WWEIA-PA}=\left(\hat{\boldsymbol{\Upsilon}}_{\mathrm{ss}}\!+\!\lambda \mathbf{I}+\frac{\mu}{K_{\rm G}}  \tilde{\boldsymbol{\Upsilon}}_{\mathrm{sg}}\right)^{-1}\mathbf{W}\mathbf{U}^H\bar{\mathbf{H}}_{\rm{ss}}.
\label{WWEIAPA}
\end{align}

Similarly, in \secref{Algorithm-A}, we also need to modify the original closed-form solution with the MMSE criterion in \eqref{20} as
\begin{align}
\mathbf{P}_{\rm MMSEIA-PA}=\beta\left(\boldsymbol{\Upsilon}_{\mathrm{ss}}+\varsigma \tilde{\boldsymbol{\Upsilon}}_{\mathrm{sg}}+ \frac{K_{\mathrm{S}}\sigma_{\rm s}^2}{P_\mathrm T}\mathbf{I}\right)^{-1} \bar{\mathbf{H}}_{\mathrm{ss}}.
\label{MMSEIAPA}
	\end{align}
    
Considering that the approximation may introduce an error that could degrade performance, we further analyze the factors influencing the approximation error.
\begin{ppn}\label{approximation}
Assuming that the terrestrial users of the single terrestrial BS located at the sub-satellite point follow a uniform distribution within the terrestrial BS radius \( R_{\mathrm{bs}} \), and the user density is \( \rho_{\mathrm{tut}} \). The $(i, j)$-th element of the squared error matrix $\mathcal{E}_{\mathrm {sg}} \in \mathbb{R}^{M_\mathrm{S} \times M_\mathrm{S}} $ between the integral-form expression \( \boldsymbol{\Upsilon}_{\mathrm{sg}}^{\rm int} \) and its approximation \( \tilde{\boldsymbol{\Upsilon}}_{\mathrm{sg}} \) can be computed as
\begin{align}
     \textstyle\left[ \mathcal{E}_{\mathrm {sg}}\right]_{i, j} =  \eta^2 \left| \frac{ \pi R_{\mathrm{bs}}^{2}} {h_{\rm{sat}}^{2}}-\int_{0}^{R_{\mathrm{bs}}}\frac{2 \pi r J_0\left(\frac{\pi r \omega}{R_{\rm{sat}}}\right) \mathrm{d} r }{h_{\rm{sat}}^{2}+r^2}\right|^2,
     \label{approx}
\end{align}
where we define $\omega=\sqrt{(m_{a}- m_{p})^2+(n_{b}- n_{q})^2}$ and \(\eta = \frac{G_{\rm T}G_{\rm R}c^2 \rho_{\mathrm{tut}}}{(4\pi f)^2 }\). The gradients satisfy the following inequalities
\begin{equation}
	\begin{aligned}
         \nabla_{R_{\rm{bs}}}  \left[ \mathcal{E}_{\mathrm {sg}}\right]_{i, j}\geq 0,\,\,  \nabla_{\rho_{\mathrm{tut}}}  \left[ \mathcal{E}_{\mathrm {sg}}\right]_{i, j}\geq 0,\,\, \nabla_{f}  \left[ \mathcal{E}_{\mathrm {sg}}\right]_{i, j}\leq 0.
	\end{aligned}
\end{equation}
\end{ppn}
\begin{pf}
According to equation \eqref{integral}, we can derive the error between the integral-form  \( \boldsymbol{\Upsilon}_{\mathrm{sg}}^{\rm int} \) and its approximation \( \tilde{\boldsymbol{\Upsilon}}_{\mathrm{sg}} \) as in \eqref{app}.
\begin{figure*}[htbp]
	\normalsize
    \begin{align}
		 \textstyle\left[ \mathcal{E}_{\mathrm {sg}}\right]_{i, j} =\left|{\left[ \tilde{\boldsymbol{\Upsilon}}_{\mathrm{sg}}\right]}_{i, j}-\left[\boldsymbol{\Upsilon}_{\mathrm{sg}}^{\rm int}\right]_{i, j}\right|^2 = \eta^2 \left|\frac{\pi R_{\mathrm{bs}}^{2}}{h_{\mathrm{sat}}^{2}}-\int_{0}^{2\pi}\!\!\int_{0}^{R_{\mathrm{bs}}}\frac{\mathrm{e}^{\frac{j \pi}{R_{\mathrm{sat}}}\left[\left(m_{a}- m_{p}\right)r \cos \phi+\left(n_{b}- n_{q}\right)r\sin \phi\right]} r \!\! \mathrm{~d}r\! \!\mathrm{~d} \phi}{h_{\mathrm{sat}}^{2}+r^{2}}\right|^2.
		 \label{app}
    \end{align}
\hrulefill
	\vspace{-5mm}
\end{figure*}

 Using the first kind of Bessel function \( \int_{0}^{2 \pi} e^{j a \sin \phi} \, \mathrm{d} \phi = 2 \pi J_{0}(a) \), the integral can be simplified as $ \left[ \mathcal{E}_{\mathrm {sg}}\right]_{i, j} =  \eta^2 \left| \frac{ \pi R_{\mathrm{bs}}^{2}} {h_{\rm{sat}}^{2}}-\int_{0}^{R_{\mathrm{bs}}}\frac{2 \pi r J_0\left(\frac{\pi r \omega}{R_{\rm{sat}}}\right) \mathrm{d} r }{h_{\rm{sat}}^{2}+r^2}\right|^2$.
The gradient of $\left[ \mathcal{E}_{\mathrm {sg}}\right]_{i, j}$ with respect to $R_{\rm bs}$ is given
    \begin{align}
		\begin{split} 
			\nabla_{R_{\rm{bs}}}  \left[ \mathcal{E}_{\mathrm {sg}}\right]_{i, j} = &\textstyle 2\eta^2 \left( 
			\frac{2\pi R_{\mathrm{bs}}}{h_{\mathrm{sat}}^{2}} - \frac{ 2 \pi R_{\rm{bs}}J_0\!\left(\frac{\pi  R_{\rm{bs}}\omega}{R_{\rm{sat}}}\right)}{h_{\mathrm{sat}}^{2}+R_{\rm{bs}}^2} \right)\\&\textstyle\times \left(\frac{ \pi R_{\mathrm{bs}}^{2}} {h_{\rm{sat}}^{2}}\!-\!\int_{0}^{R_{\mathrm{bs}}}\frac{2 \pi r J_0\left(\!\frac{\pi r \omega}{R_{\rm{sat}}}\!\right) \!\mathrm{d} r }{h_{\rm{sat}}^{2}+r^2}\!\right),
         \label{grad}       
		\end{split}
        \end{align} 
where we can deduce that 
\begin{align}
		\textstyle\frac{ 2 \pi R_{\rm{bs}}J_0\left(\frac{\pi  R_{\rm{bs}}\omega}{R_{\rm{sat}}}\right)}{h_{\mathrm{sat}}^{2}+R_{\rm{bs}}^2} \!\leq\! \frac{2\pi R_{\mathrm{bs}}}{h_{\mathrm{sat}}^{2}},\ \textstyle \int_{0}^{R_{\mathrm{bs}}}\frac{2 \pi r J_0\left(\frac{\pi r \omega}{R_{\rm{sat}}}\right) \mathrm{d} r }{h_{\rm{sat}}^{2}+r^2}
    \!\leq \!  \frac{ \pi R_{\mathrm{bs}}^{2}} {h_{\rm{sat}}^{2}}.
        \end{align} 

    Given that \(\eta = \frac{G_{\rm T}G_{\rm R} c^2 \rho_{\mathrm{tut}}}{(4\pi f)^2}\), we can derive the gradients inequalities:
     $\nabla_{R_{\rm{bs}}}  \left[ \mathcal{E}_{\mathrm {sg}}\right]_{i, j}\geq 0$, $\nabla_{\rho_{\mathrm{tut}}}  \left[ \mathcal{E}_{\mathrm {sg}}\right]_{i, j}\geq 0$, and $ \nabla_{f}  \left[ \mathcal{E}_{\mathrm {sg}}\right]_{i, j}\leq 0$.
\end{pf}
\begin{figure}[H]
\centering
\includegraphics[width=0.84\linewidth,trim=3cm 9.2cm 3.7cm 9.2cm,clip]{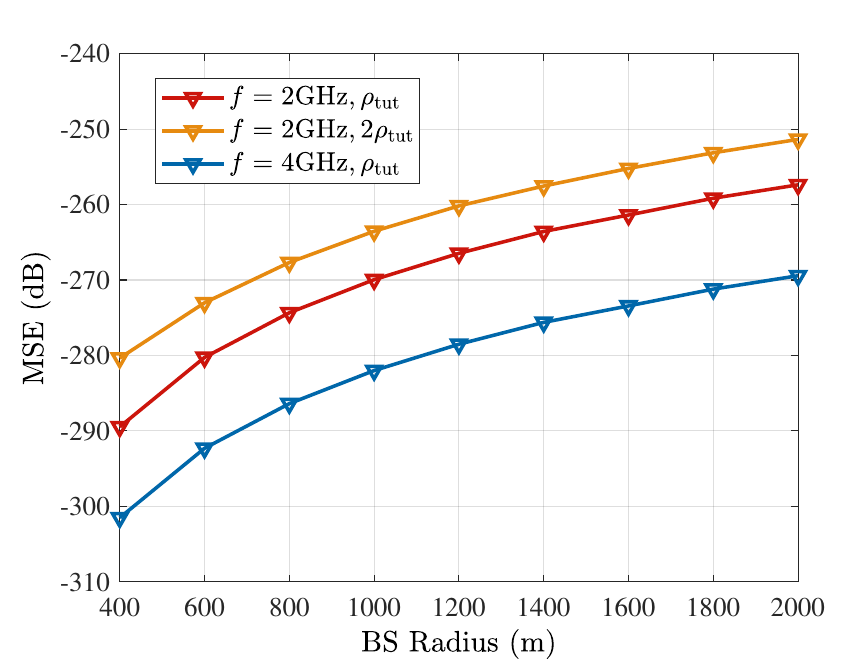}
		\caption{The MSE between $\boldsymbol{\Upsilon}_{\mathrm{sg}}^{\rm int}$ and $\tilde{\boldsymbol{\Upsilon}}_{\mathrm{sg}}$ vs $R_{\rm{bs}}$, $\rho_{\mathrm{tut}}=0.0001/\mathrm{m}^2$.}
	\label{pic5-MSE}	
\end{figure}

In addition, the simulation results presented in \figref{pic5-MSE} validate the monotonic relationship between the terrestrial BS radius \( R_{\rm{bs}} \) and the approximation error under varying carrier frequencies \( f \) and user densities \( \rho_{\rm{tut}} \). Specifically, we simulated the MSE of the approximate term for the single terrestrial BS located at the sub-satellite point. 
The results demonstrate that the MSE increases with the growth of the terrestrial BS radius \( R_{\rm{bs}} \) and user density \( \rho_{\rm{tut}} \), while it decreases as the carrier frequency \( f \) increases. 

 \begin{remark}
This approximation refers to that when the position of the terrestrial UTs and the corresponding terrestrial BS exactly coincide, we have $\lim_{R_{\rm bs} \to 0}  \left[ \mathcal{E}_{\mathrm {sg}}\right]_{i, j}=0$ and the approximate term \( \tilde{\boldsymbol{\Upsilon}}_{\mathrm{sg}} \) is identical to \({\boldsymbol{\Upsilon}}_{\mathrm{sg}}^{\rm int} \). 
 \end{remark}

\begin{remark}
The computational complexity associated with the term \(\boldsymbol{\Upsilon}_{\mathrm{sg}}^\mathrm{int}\) is significantly reduced when approximated by \(\tilde{\boldsymbol{\Upsilon}}_{\mathrm{sg}}\), with the complexity order decreasing from \(\mathcal{O}\left(M_{\mathrm{S}}^{2}N_{\rm G} N_r N_{\phi}\right)\) to \(\mathcal{O}\left(M_{\mathrm{S}}^{2}N_{\rm G}\right)\).
 \end{remark}

\section{Numerical Results}\label{Simulation}
In this section, we employ the Monte Carlo method to evaluate the proposed schemes. We consider a sub-6GHz LEO satellite communication system, where the antenna elements are spaced at half-wavelength intervals along both the $x$ and $y$ axes \cite{9110855}. \textcolor{black}{During the simulation process, we focus on the LoS terrestrial UTs, which typically suffer stronger satellite interference and set the channel Rician factor to be $\kappa_k=\kappa_{\rm s}$ for all UTs.} The terrestrial BS has a coverage radius of $500$ meters. We calculate
$\mathrm{SNR} = P_{\rm T} - 10\log K_{\rm S} + G_{\rm T} + 10\log M_{\rm S} - \mathrm{PL} + G_{\rm R} - 10\log (k_{\rm bol} [T+({F}-1)T_0] B)$
where \( k_{\rm bol} \), $T_0$ and \( B \) are the Boltzmann constant, standard temperature, and system bandwidth. The noise power \( \sigma_{\rm s}^2 \) varies with the $\rm SNR$. The remaining satellite system configurations are shown in Tab. \ref{tab:1}.
	\begin{table}[!t]
		\centering
		\caption{Satellite System Parameters \cite{3GPPTR38.811,10440321,9815078}}
		\label{tab:1}
		\small
		\begin{tabular}{cc}
			\toprule
			Parameter  &  Value  \\
			\midrule
			Orbit Altitude $h_{\rm sat}$ & $600 \ \rm{km}$ \\
			Satellite Coverage Radius  $R_{\rm sat}$ & $630 \ \rm km$\\
			Carrier Frequency $f$& $2\ \rm{GHz}$\\
            Number of Antennas $M_{\rm S}$&$8 \times 8$\\
	        Rician Factor $\kappa_{\rm s}$ & $10 \ \rm{dB}$\\
            Satellite Transmit Power $P_{\rm T}$ & $25 \ \rm{dBW}$ \\
  	 Per-Antenna Gain $ G_{\rm T}$, $G_{\rm R}$ & $6 \ \rm{dBi}$, $ 0\ \rm{dBi}$ \\
           Noise Figure $F$ & $9 \ \rm {dB}$\\
          Noise Temperature $T$ & $290 \ \rm {K}$\\
			\bottomrule
		\end{tabular}
	\end{table}

\subsection{Performance Comparison of Beamforming Schemes}
In this subsection, we validate the effectiveness of our \textcolor{black}{TX} beamforming algorithms and compare the following schemes:
	\begin{itemize}
		\item `\textbf{MRT}'/`\textbf{ZF}'/`\textbf{MMSE}'/`\textbf{WMMSE}' (Baseline): Four conventional linear \textcolor{black}{TX} beamforming \cite{1494410,1391204,4712693}. 
		\item `\textbf{WQTIA}': Robust \textcolor{black}{TX} beamforming based on MCQT for inter-system interference avoidance (IA) in Algorithm \ref{A1}.
		\item `\textbf{WWEIA}': Robust \textcolor{black}{TX} beamforming based on WWE for inter-system IA in Algorithm \ref{A2}.
        \item `\textbf{MMSEIA}': Closed-form robust \textcolor{black}{TX} beamforming based on MMSE for inter-system IA in Algorithm \ref{A3}.
	\end{itemize}
    
\begin{figure}[!t]
	\centering
\includegraphics[width=1\linewidth,trim=0.5cm 0.1cm 1.5cm 0.1cm,clip] {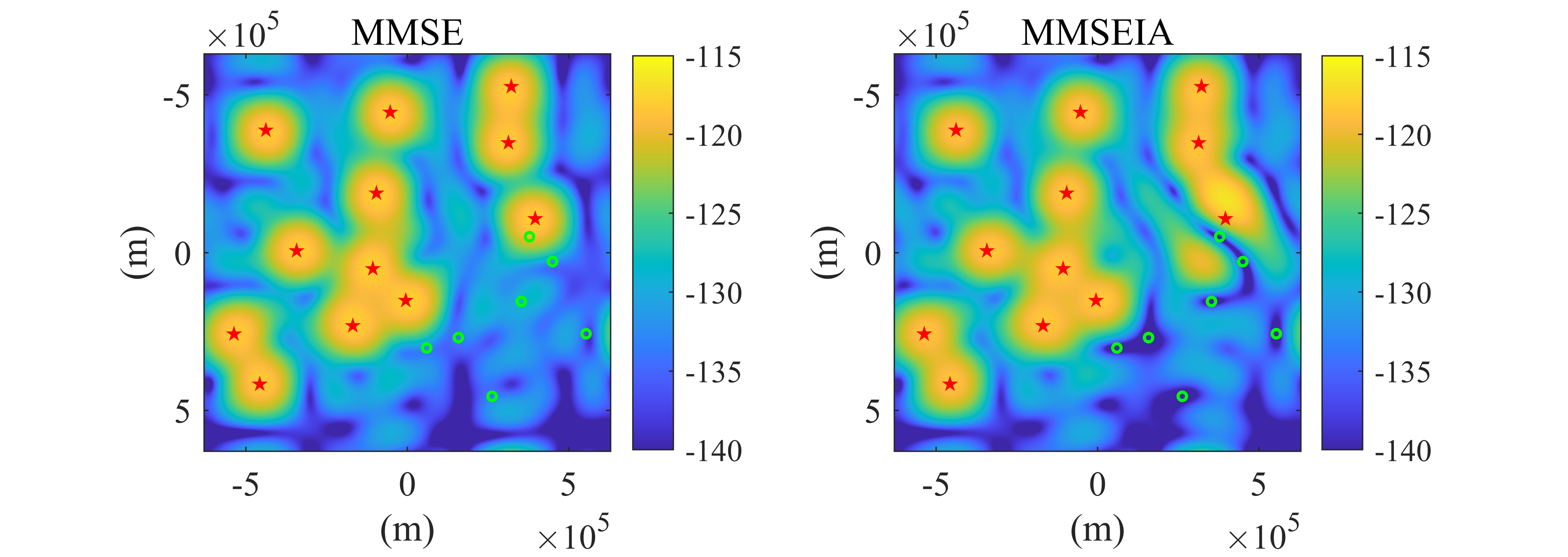}
\caption{Satellite beam patterns (red stars represent satellite UTs and green circles represent terrestrial UTs), $I_\mathrm{thr}=-150 \ \mathrm{dBW}$, $\mathrm{SNR}=10\ \mathrm{dB}$.}
	\label{1-pattern}	
\end{figure}

In \figref{1-pattern}, we present the beam patterns generated by different \textcolor{black}{TX} beamforming schemes. A comparison of these schemes reveals that the interference power at the positions of terrestrial UTs (indicated by green circles) is lower with the `MMSEIA' scheme than with the `MMSE' scheme, demonstrating the intuitive effect of interference mitigation.

\begin{figure}[!t]
 		\centering	\includegraphics[width=0.94\linewidth,trim=0.01cm 9.5cm 0.01cm 9.3cm,clip]{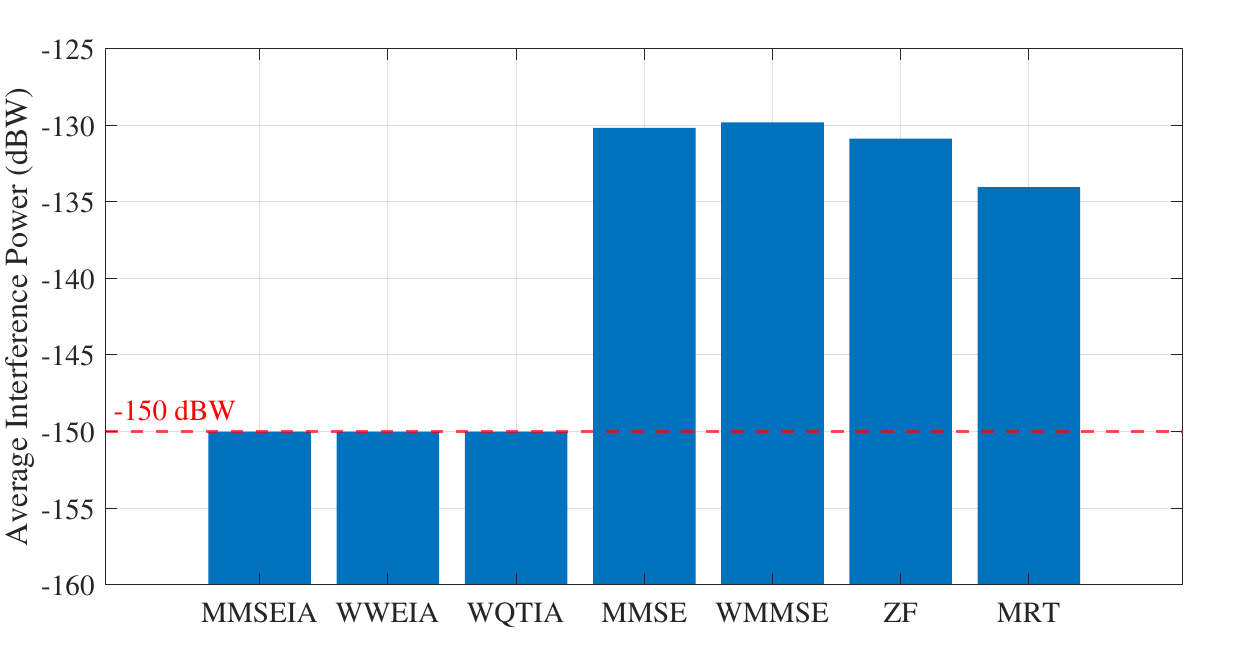}
\label{1_avginf_snr=10dB}
    \caption{Average interference power, $\mathrm{SNR}=10\ \mathrm{dB}$.}
    	\label{1-avginf}
 \end{figure}
\begin{figure}[!t]
		\centering
\includegraphics[width=0.8\linewidth,trim=3.5cm 9.3cm 3.5cm 9.5cm,clip]{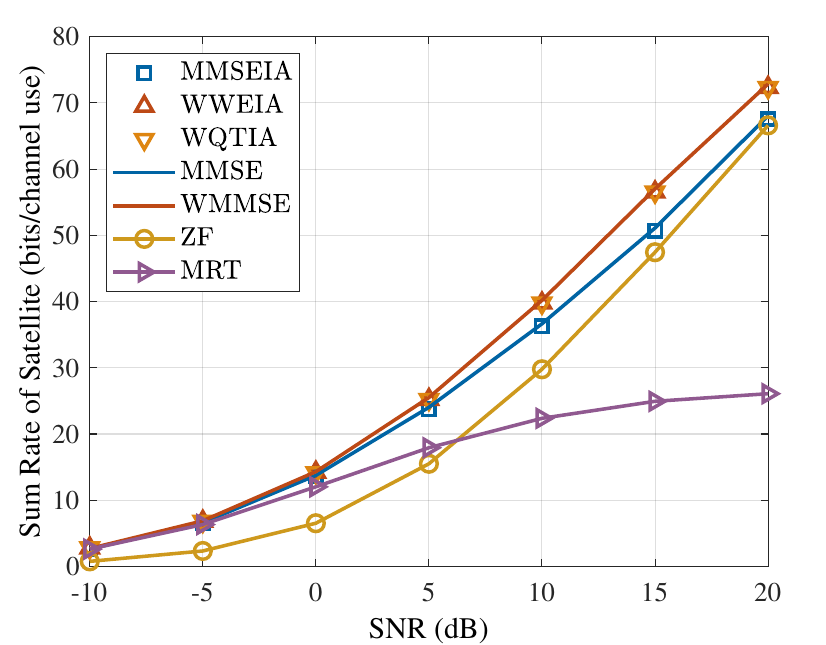}		\caption{Satellite sum rate vs $\mathrm{SNR}$, $I_\mathrm{thr}=-150\ \rm{dBW}$.}
		\label{1-sumrate}	
	\end{figure}
\figref{1-avginf} and \figref{1-sumrate} illustrate the comparison of the average interference power performance and sum rate of the satellite for a UPA configuration with \( M_\mathrm{S} = 8 \times 8 \), satellite UT number \( K_\mathrm{S} = 12 \), terrestrial BS number \( N_\mathrm{G} = 7 \), terrestrial UT number per BS \( \bar K_\mathrm{G} = 10 \), and an average interference threshold of \( I_\mathrm{thr} = -150 \ \mathrm{dBW} \).
In \figref{1-avginf}, the three \textcolor{black}{TX} beamforming schemes `MMSEIA', `WWEIA', and `WQTIA' successfully achieve the interference threshold of $-150 \ \rm{dBW}$ at $\rm SNR=10 \ \rm{dB}$. This performance illustrates their effectiveness in interference management.
In \figref{1-sumrate}, the `MMSEIA' scheme shows a sum rate close to that of the `MMSE' scheme, indicating its effectiveness in sustaining capacity. Furthermore, both the `WWEIA' and `WQTIA' schemes outperform `MMSE' and closely approach the performance of `WMMSE'. These two iterative algorithms significantly enhance the sum rate, while the closed-form `MMSEIA' operates with lower computational complexity.

\begin{figure}[!t]
  \centering
\includegraphics[width=0.78\linewidth,trim=3cm 9.2cm 4cm 9.5cm,clip]{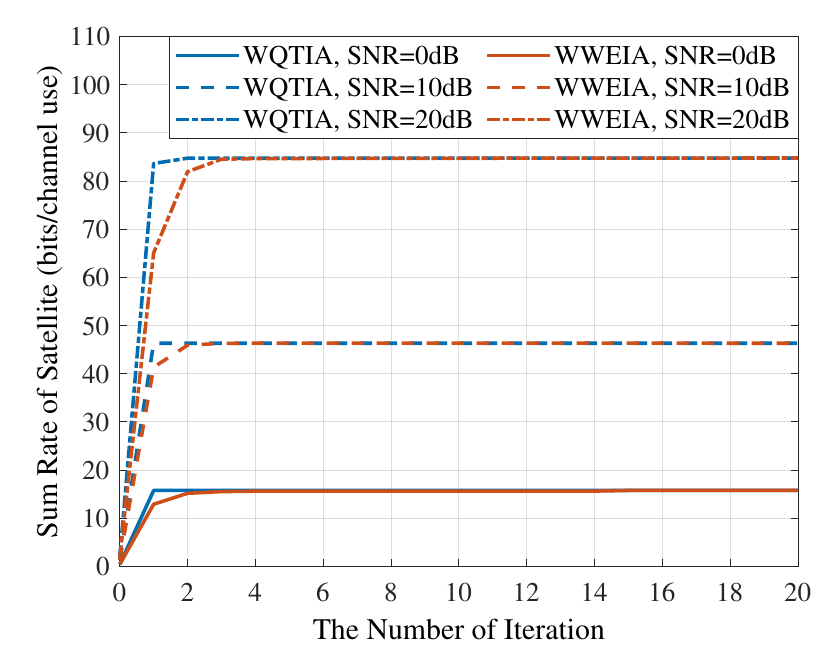}
\caption{The convergence of satellite sum rate.}
 \label{1-iter_sumrate}	
 \end{figure}

 For the iterative algorithms `WWEIA' and `WQTIA', we study their convergence behaviors in satellite sum rate. As illustrated in \figref{1-iter_sumrate}, the satellite sum rate performances of the proposed schemes `WWEIA' and `WQTIA' quickly converge within the specified number of iterations. The results are verified at $\rm{SNR}=0 \ \rm{dB}$, $ \rm{SNR}=10 \ \rm{dB}$, and $ \rm{SNR}=20 \ \rm{dB}$, demonstrating consistent convergenc performance. We also observe that `WWEIA' takes more iterations to converge as $\rm SNR$ increases.  
\subsection{Effect of Interference Thresholds and Satellite UT Numbers}
\begin{figure}[!t]
\centering
\includegraphics[width=0.78\linewidth,trim=3cm 9.3cm 3.5cm 9.4cm,clip]{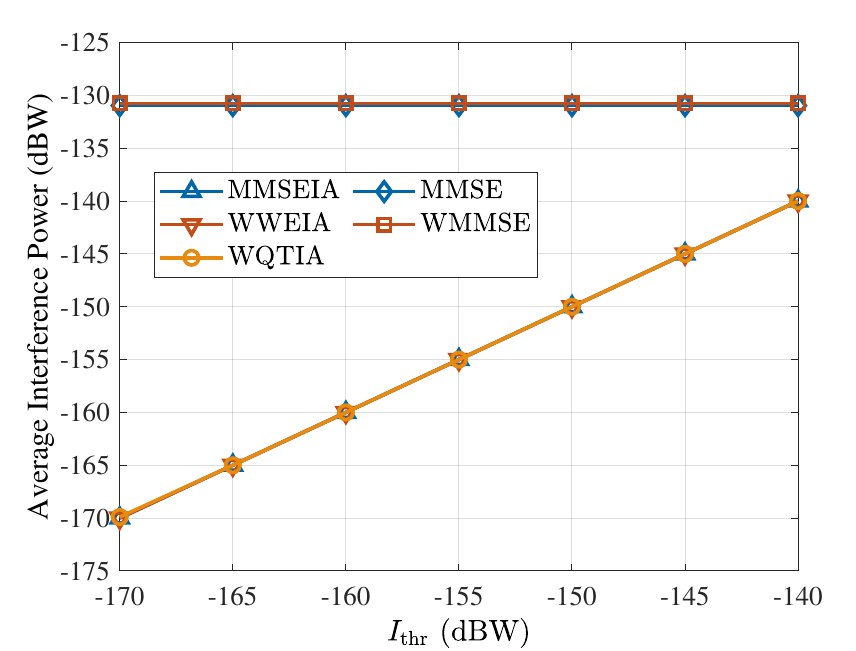}
\caption{Average interference vs $I_{\rm{thr}}$, $\mathrm{SNR}=10\ \mathrm{dB}$.}
\label{sim2_Ithr_avginf}
\end{figure}

\begin{figure}[!t]
	\centering	\includegraphics[width=0.78\linewidth,trim=3cm 9.3cm 3.7cm 9.2cm, clip]{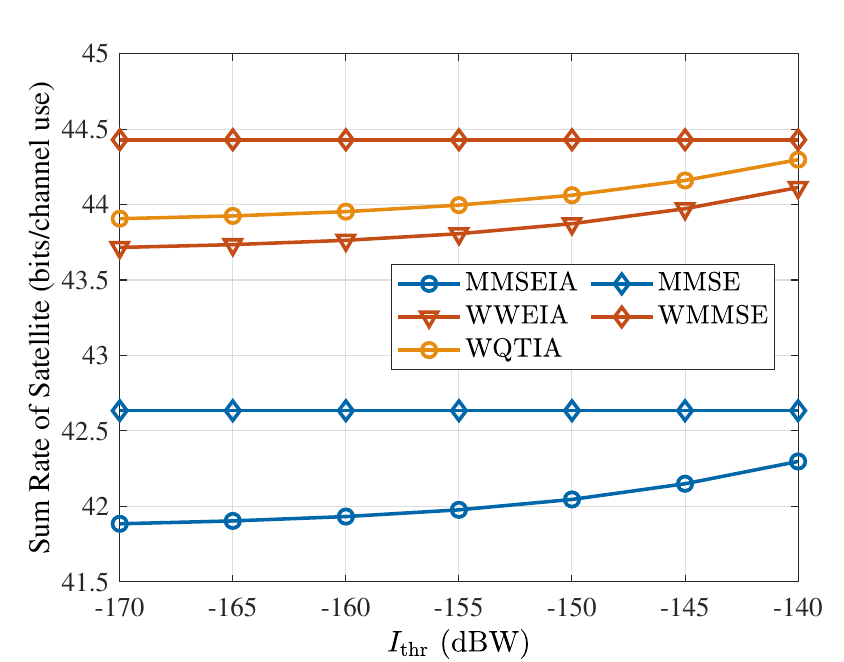}	\caption{Satellite sum rate vs $I_{\rm{thr}}$, $\mathrm{SNR}=10\ \mathrm{dB}$.}
\label{sim2_Ithr_sumrate}	
\end{figure}

  \figref{sim2_Ithr_avginf} and \figref{sim2_Ithr_sumrate} illustrate the impact of interference thresholds on the satellite sum rate and average interference power performance. 
   As illustrated in \figref{sim2_Ithr_avginf}, the average interference power remains the same with `MMSE' and `WMMSE' schemes, but decreases adaptively as the interference threshold \(I_{\mathrm{thr}}\) is reduced with `MMSEIA', `WWEIA', and `WQTIA' schemes. As shown in \figref{sim2_Ithr_sumrate}, when we set the $\rm SNR$ as $10 \ \rm{dB}$, the sum rate for the `MMSE' and `WMMSE' schemes remains unchanged. However, the satellite sum rate for both the `MMSEIA' and `WWEIA' schemes decreases slightly as the interference threshold \( I_{\mathrm{thr}} \) is reduced. Specifically, when \( I_{\mathrm{thr}} = -170\ {\rm dBW} \), all three \textcolor{black}{TX} beamforming algorithms exhibit a sum rate reduction of approximately \( 1\ \% \) compared to the case when \( I_{\mathrm{thr}} = -140\ {\rm dBW} \).  
\begin{figure}[!t]
	\centering
\includegraphics[width=0.78\linewidth,trim=3.5cm 9.4cm 3.5cm 9.7cm,clip]{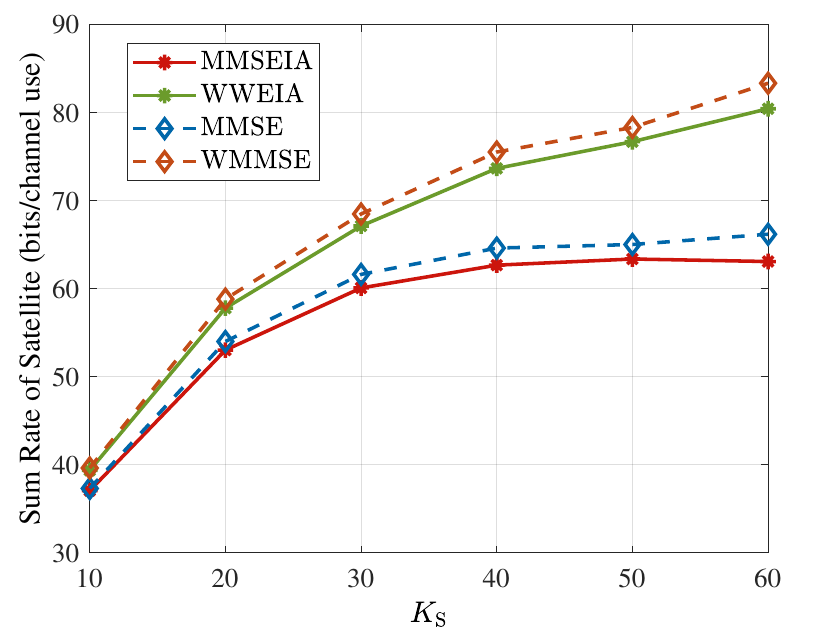}
\caption{Satellite sum rate vs $K_{\rm{S}}$, $I_\mathrm{thr} = -150 \ \mathrm{dBW}$, $\mathrm{SNR}=10\ \mathrm{dB}$.}
\label{sim3_Ks_sumrate}	
\end{figure} 

Furthermore, we study the impact of different satellite UT numbers on the proposed \textcolor{black}{TX} beamforming schemes, with a fixed number of terrestrial BSs and a fixed number of terrestrial UTs. Taking the lower-complexity `MMSEIA' and `WWEIA' as examples, it can be observed in \figref{sim3_Ks_sumrate} that the sum rate of the satellite increases with the number of satellite UTs \( K_{\rm{S}} \). However, the rate of increase diminishes as \( K_{\rm{S}} \) grows, and when \( K_{\rm{S}} \geq 40 \), the sum rate of `MMSEIA' approaches a plateau, remaining nearly constant thereafter. Notably, the satellite sum rate performance of `MMSEIA' approaches that of `MMSE', and `WWEIA' approaches the performance of `WMMSE'.

\subsection{Validation of Position-Aided Interference Approximation}
In this subsection, we further validate the performance after interference approximation in beamforming design. The integral-form interference channel term \(\boldsymbol{\Upsilon}_{\mathrm{sg}}^{\rm int}\) is approximated by the corresponding satellite-to-terrestrial BS channels, denoted as \(\tilde{\boldsymbol{\Upsilon}}_{\mathrm{sg}}\). We fix the terrestrial BS radius at \( R_{\rm{bs}} = 500\ \mathrm{m} \) and compare the performance of the approximated schemes `MMSEIA-PA', `WWEIA-PA', and `WQTIA-PA' with the original \textcolor{black}{TX} beamforming schemes `MMSEIA', `WWEIA', and `WQTIA'.

 \begin{figure}[!t]
 		\centering
\includegraphics[width=0.94\linewidth,trim=0cm 9.3cm 0.7cm 9.5cm,clip]{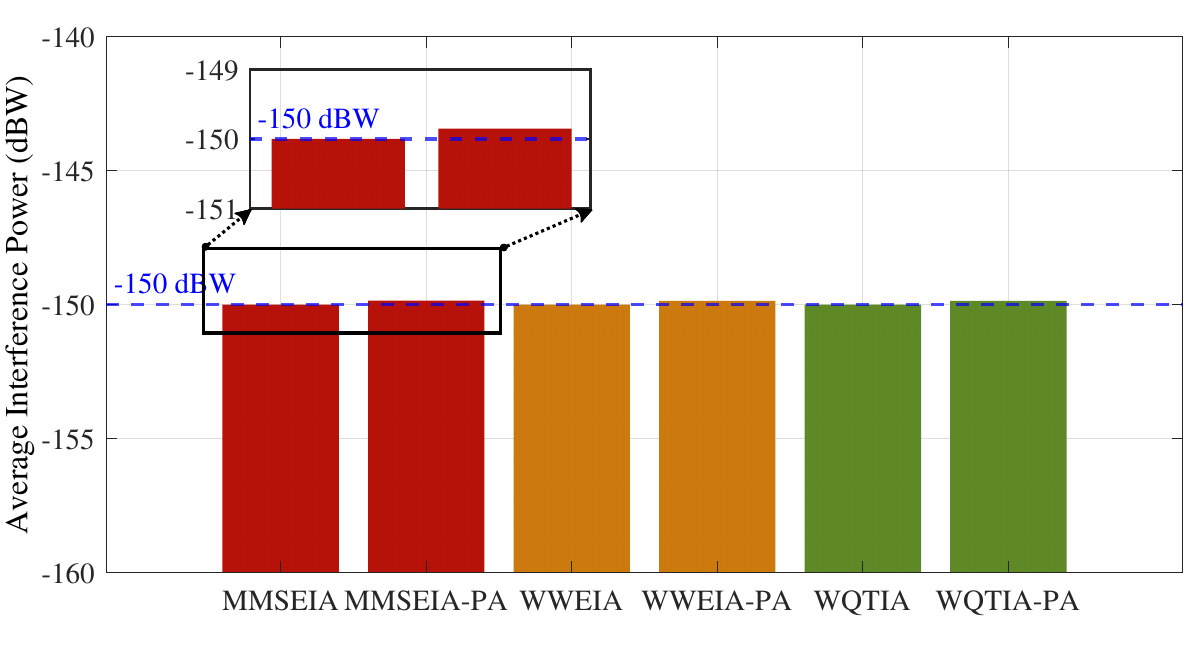}
    \caption{Average interference power, $I_\mathrm{thr}=-150\ \rm{dBW}$, $\mathrm{SNR}=10\ \mathrm{dB}$.}	\label{sim5_PA_avginf}
 \end{figure}
\begin{figure}[!t]
	\centering
\includegraphics[width=0.78\linewidth,trim=3.5cm 9.2cm 3.5cm 9.5cm,clip]{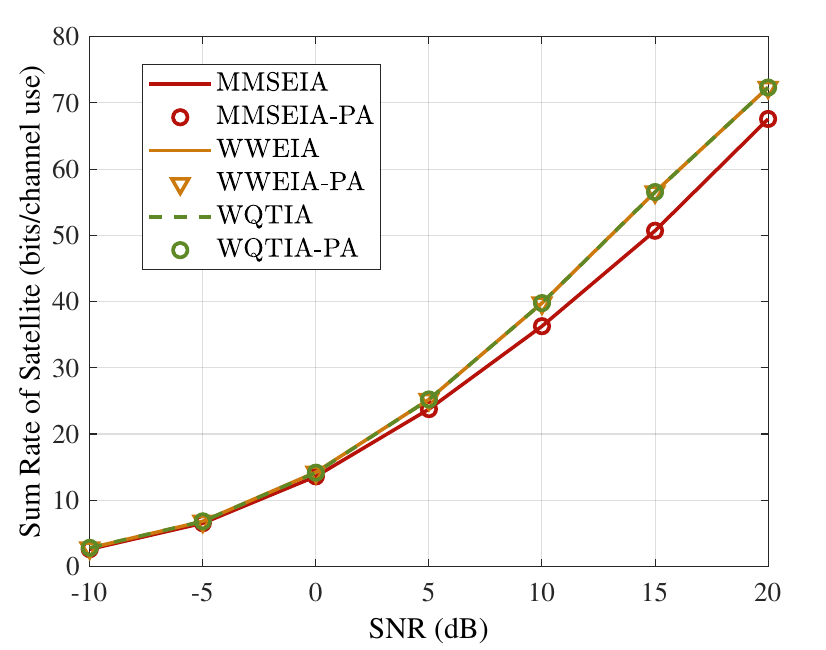}
	\caption{Satellite sum rate vs $\mathrm{SNR}$, $I_\mathrm{thr}=-150\ \rm{dBW}$.}
\label{sim5_PA_sumrate}	
\end{figure}

In \figref{sim5_PA_avginf}, the results reveal that all three PA schemes nearly meet the interference threshold of $-150 \ \rm{dBW}$ at $10 \ \rm{dB}$, representing the effectiveness of the approximation in terms of interference management.
 In \figref{sim5_PA_sumrate}, the sum rate performance of the approximated schemes is consistent with that of the original schemes.
The proposed PA method reduces computational complexity while maintaining effective interference management and high satellite sum rate performance, highlighting the practical benefits of PA schemes in ITSN systems.
 
\subsection{Comparison of Computational Complexity}
In this subsection, we compare the order of computational complexity of the schemes considered in Tab. \ref{complexity}. Here, \(N_A\) and \(N_B\) represent the number of outer loop iterations of `WQTIA' and `WWEIA`, respectively. From this comparison, we can conclude that `WWEIA' significantly reduces the complexity compared to `WQTIA', primarily due to its more efficient iterative procedures. The closed-form solution, `MMSEIA', exhibits a lower complexity than both `WQTIA' and `WWEIA'. Furthermore, PA schemes reduce the complexity by decreasing the dimension from the number of discrete samples \( N_{\rm{G}}N_r N_{\phi} \) to the number of terrestrial BSs \( N_{\rm{G}} \).
    \begin{table}[!t]
		\centering
	\caption{Complexity Comparison}
		\label{complexity}
		\footnotesize
		\begin{tabular}{cc}
			\toprule
			$\mathbf{Scheme}$  &  $\mathbf{Complexity  \ Order}$  \\
			\midrule
            WQTIA & \(\mathcal{O}\left(N_A\left(T_A \left(M_{\rm S}^3 K_{\rm S}^{3}\right)+M_{\mathrm{S}}^{2}N_{\rm G} N_r N_{\phi}\right)\right)\) \\
		WWEIA  &\!\!\!\! \(\mathcal{O}\!\left(N_B\!\left(T_B + M_{\rm S}^2K_{\rm S} + M_{\mathrm{S}}^{2} N_{\rm G} N_r N_{\phi}+M_{\rm S}^3 \right)\right)\) \\
            MMSEIA  & \(\mathcal{O}\left(M_{\mathrm{S}}^{2} N_{\rm G} N_r N_{\phi} + M_{\mathrm{S}}^{3}\right)\) \\
			\midrule
            WQTIA-PA & \(\mathcal{O}\left(N_A\left(T_A \left(M_{\rm S}^3 K_{\rm S}^{3}\right)+M_{\mathrm{S}}^{2}N_{\rm G}\right)\right)\) \\
			WWEIA-PA & \!\!\!\!\(\mathcal{O}\!\left(N_B\!\left(T_B + M_{\rm S}^2K_{\rm S}^2 + M_{\mathrm{S}}^{2} N_{\rm G}+M_{\rm S}^3 \right)\right)\)\\
			MMSEIA-PA &  \(\mathcal{O}\left( M_{\mathrm{S}}^{2} N_{\mathrm{G}} + M_{\mathrm{S}}^{3}\right)\) \\
			\bottomrule
		\end{tabular}
\vspace{-5mm}
\end{table}

\section{Conclusion}\label{Conclusion}
This paper investigated robust \textcolor{black}{TX} beamforming based on sCSI, \textcolor{black}{mitigating} satellite-to-terrestrial UT interference \textcolor{black}{in ITSNs}. We established an integral-form interference model free of shared CSI and designed robust \textcolor{black}{TX} beamforming schemes under the power budget and the interference threshold. The WSR problem was iteratively solved after MCQT, while an equivalent WMMSE framework achieved a lower-complexity iterative optimization. Furthermore, we derived a closed-form solution for the MMSE criterion and applied a bisection method to meet the interference threshold. To reduce reliance on complex integral calculations, we proposed an approximation scheme based on terrestrial BS positions. Simulation results showed \textcolor{black}{the level of interference mitigation that can be achieved with the proposed schemes and verified the effectiveness of the PA schemes, allowing efficient spectrum sharing.} In future work, we will extend these schemes to satellite uplinks and terrestrial BS transmission within the ITSN framework.

	
	%

	
	
\appendices
\section{Proof of Proposition \ref{exp}} \label{app_exp} 
	We substitute \(\mathbf{h}_{\mathrm{ss},k} \) with ${\mathbf{b}}_k=\mathrm{e}^{-j \psi_k}\mathbf{h}_{\mathrm{ss},k}$ and \(\psi_k=2 \pi\left(t v_k^{\rm{sat}}\!-\! f \tau_{k}^{\rm{min}}\right)\) in the sum rate of the satellite system as follows
\begin{equation}
\begin{split}
&  \sum_{k=1}^{K_\mathrm{S}} \mathbb{E}_{\mathbf{H}_{\rm ss}}\left\{a_k\log_{2}\left(1 + \frac{\mathbf{p}_{k}^{H} \mathbf{h}_{\mathrm{ss},k} \mathbf{h}_{\mathrm{ss},k}^H \mathbf{p}_{k}}{\sum_{i \neq k} \mathbf{p}_{i}^{H} \mathbf{h}_{\mathrm{ss},k} \mathbf{h}_{\mathrm{ss},k}^H \mathbf{p}_{i} + \sigma_k^{2}}\right)\right\} \\
&=\! \sum_{k=1}^{K_\mathrm{S}} \mathbb{E}_{\mathbf{B}}\left\{\!a_k\log_{2}\left(\! 1 + \frac{\mathbf{p}_{k}^{H} \mathrm{e}^{j \psi_k}{\mathbf{b}}_{k} \mathrm{e}^{-j \psi_k}{\mathbf{b}}_{k}^H \mathbf{p}_{k}}{\sum_{i \neq k} \mathbf{p}_{i}^{H}  \mathrm{e}^{j \psi_k}{\mathbf{b}}_{k} \mathrm{e}^{-j \psi_k}{\mathbf{b}}_{k}^H \mathbf{p}_{i} +\sigma_k^{2}}\!\right)\!\right\}\\
&  = \sum_{k=1}^{K_\mathrm{S}} \mathbb{E}_{\mathbf{B}}\left\{a_k\log_{2}\left(1+ \frac{\mathbf{p}_{k}^{H} {\mathbf{b}}_{k} {\mathbf{b}}_{k}^H \mathbf{p}_{k}}{\sum_{i \neq k} \mathbf{p}_{i}^{H} {\mathbf{b}}_{k}{\mathbf{b}}_{k}^H \mathbf{p}_{i} +\sigma_k^{2}}\right)\right\}.
\label{phase}
\end{split}
\end{equation}
This completes the proof.

\section{Proof of Proposition \ref{closed-form}} \label{app_closed-form} 
The Lagrangian function is derived as
	\begin{align}
	&\mathcal{L}(\mathbf{P}, \beta, \lambda)=\frac{\operatorname{Tr}\left\{ \mathbf{P}^{H}\boldsymbol{\Upsilon}_{\mathrm{ss}}\mathbf{P} \right\}}{\beta^{2}} - \frac{\operatorname{Tr}\left\{\bar{\mathbf{H}}_\mathrm{s s}^H \mathbf{P}\right\}}{\beta}-\frac{\operatorname{Tr}\left\{\mathbf{P}^H \bar{\mathbf{H}}_\mathrm{ss}\right\}}{\beta}\notag\\
    & + K_{\mathrm{S}} + \frac{K_{\mathrm{S}} \sigma_{\rm s}^{2}}{\beta^{2}} + \frac{\varsigma }{\beta^2} \operatorname{Tr}\left\{ \mathbf{P}^{H}\boldsymbol{\Upsilon}_{\mathrm{sg}}\mathbf{P} \right\} + \lambda \left(\operatorname{Tr}\left\{\mathbf{P} \mathbf{P}^{H}\right\} - P_{\mathrm{T}}\right).
    \label{13}
	\end{align}
Then, we set the gradient of the Lagrange function to zero
\begin{align}
	\nabla_{\mathbf{P}^*} \mathcal{L}=\frac{\boldsymbol{\Upsilon}_{\mathrm{ss}} \mathbf{P}}{\beta^{2}}-\frac{\bar{\mathbf{H}}_\mathrm{s s}}{\beta}+\frac{\varsigma }{\beta^2}{\boldsymbol{\Upsilon}_{\mathrm{sg}}\mathbf{P}}+\lambda \mathbf{P}=\mathbf{0},
        	\label{14}
\end{align}
and we obtain
\begin{align}
\mathbf{P}=\beta\left(\boldsymbol{\Upsilon}_{\mathrm{ss}}+\varsigma \boldsymbol{\Upsilon}_{\mathrm{sg}}+\lambda \beta^{2} \mathbf{I}\right)^{-1} \bar{\mathbf{H}}_\mathrm{s s}.
	\label{15}
\end{align}
We define $\zeta=\lambda\beta^2$, $\mathbf{A} = \boldsymbol{\Upsilon}_{\mathrm{ss}} + \varsigma \boldsymbol{\Upsilon}_{\mathrm{sg}} + \zeta \mathbf{I}$, $\mathbf{P} = \beta \mathbf{A}^{-1}\bar{\mathbf{H}}_\mathrm{s s}$ and $\bar {\boldsymbol{\Upsilon}}_{\mathrm{ss}}=\bar{\mathbf{H}}_\mathrm{s s}\bar{\mathbf{H}}_\mathrm{s s}^H$. Through normalization, we have $\beta = \sqrt{\frac{P_\mathrm{T}}{\operatorname{Tr}\left\{\mathbf{A}^{-2}\bar{\boldsymbol{\Upsilon}}_{\mathrm{ss}}\right\}}}$, and the problem is transformed into an unconstrained optimization problem: $\min_{\zeta} f(\mathbf{P}(\zeta), \beta(\zeta))$, where the objective function is
	\begin{equation}
		\begin{split}
	f(\zeta)=&\operatorname{Tr}\left\{ \mathbf{A}^{-1} \boldsymbol{\Upsilon}_{\mathrm{ss}}\mathbf{A}^{-1} \bar {\boldsymbol{\Upsilon}}_{\mathrm{ss}}\right\}-2\operatorname{Tr}\left\{\mathbf{A}^{-1}\bar {\boldsymbol{\Upsilon}}_{\mathrm{ss}}\right\}
    +K_{\mathrm{S}}\\&+\frac{K_{\mathrm{S}}\sigma_{\rm s}^{2}}{P_\mathrm{T}} \operatorname{Tr}\left\{\mathbf{A}^{-2} \bar {\boldsymbol{\Upsilon}}_{\mathrm{ss}}\right\}+\varsigma  \operatorname{Tr}\left\{\boldsymbol{\Upsilon}_{\mathrm{sg}}\mathbf{A}^{-1} \bar {\boldsymbol{\Upsilon}}_{\mathrm{ss}}\mathbf{A}^{-1} \right\}.
		\end{split}
		\label{18}
	\end{equation}
We calculate the gradient of \eqref{13} with respect to \( \zeta \) as
$\nabla_{\zeta} f
=2\operatorname{Tr}\left\{\left(\zeta-\frac{K_{\mathrm{S}} \sigma_{\rm s}^{2}}{P_\mathrm T}\right)\left(\mathbf{A}^{-3}\bar {\boldsymbol{\Upsilon}}_{\mathrm{ss}}\right)\right\}$, and set it to zero.
Therefore, we have $\zeta = \lambda \beta^2 = \frac{K_{\mathrm{S}} \sigma_{\rm s}^2}{P_\mathrm{T}}$. Substituting this into $\mathbf{P}$ and $\beta$ gives the closed-form solution
\begin{align}
&\mathbf{P}^{\star}=\beta^{\star}\left(\boldsymbol{\Upsilon}_{\mathrm{ss}}+\varsigma \boldsymbol{\Upsilon}_{\mathrm{sg}}+ \frac{K_{\mathrm{S}}\sigma_{\rm s}^2}{P_\mathrm T}\mathbf{I}\right)^{-1} \bar{\mathbf{H}}_\mathrm{s s},\\
&\textstyle\beta^{\star} = \sqrt{\frac{P_\mathrm{T}}{\left\|{\left(\boldsymbol{\Upsilon}_{\mathrm{ss}}+\varsigma \boldsymbol{\Upsilon}^{\mathrm {int}}_{\mathrm{sg}}+ \frac{K_{\mathrm{S}}\sigma_{\rm s}^2}{P_\mathrm T}\mathbf{I}\right)^{-1}\bar{\mathbf{H}}_\mathrm{s s}}\right\|_F^2}}.
\end{align}

This completes the proof.

\vspace{-1mm}
\section{Proof of Proposition \ref{dI-dk}} \label{app_dI-dk}
\vspace{-1mm}

 From \eqref{20} and \eqref{22}, we simplfy \(\mathbf{P}(\varsigma ) = \beta \mathbf{A}^{-1}\bar{\mathbf{H}}_\mathrm{s s}\), \(\alpha = \operatorname{Tr}\left\{\mathbf{A}^{-2}\bar {\boldsymbol{\Upsilon}}_{\mathrm{ss}}\right\}\) and $\bar {\boldsymbol{\Upsilon}}_{\mathrm{ss}}=\bar{\mathbf{H}}_\mathrm{s s}\bar{\mathbf{H}}_\mathrm{s s}^H$. The gradient can be derived as
\begin{align}
\begin{split}
\nabla_{\varsigma } I_\mathrm{sg} =& \frac{2\beta^2}{\alpha} \operatorname{Tr} \left\{ \bar {\boldsymbol{\Upsilon}}_{\mathrm{ss}}\mathbf{A}^{-1} \boldsymbol{\Upsilon}^{\mathrm {int}}_{\mathrm{sg}}\mathbf{A}^{-2} \right.\\& \left.\left( \mathbf{A}^{-1} \boldsymbol{\Upsilon}^{\mathrm {int}}_{\mathrm{sg}} \bar {\boldsymbol{\Upsilon}}_{\mathrm{ss}}\mathbf{A}^{-1}-\alpha \boldsymbol{\Upsilon}^{\mathrm {int}}_{\mathrm{sg}} \right) \right\},
\end{split}
\label{a4}
\end{align}
where both the complex matrix \(\mathbf{M} = \frac{2\beta^2}{\alpha} \bar {\boldsymbol{\Upsilon}}_{\mathrm{ss}} \mathbf{A}^{-1} \boldsymbol{\Upsilon}^{\mathrm {int}}_{\mathrm{sg}} \mathbf{A}^{-2}\) and matrix \(\mathbf{N} = \mathbf{A}^{-1} \boldsymbol{\Upsilon}^{\mathrm {int}}_{\mathrm{sg}} \bar {\boldsymbol{\Upsilon}}_{\mathrm{ss}}\mathbf{A}^{-1} - \alpha \boldsymbol{\Upsilon}^{\mathrm {int}}_{\mathrm{sg}}\) are Hermitian matrices.
From \textit{Von Neumann Trace Inequality}, we derive that
\begin{align}
\begin{split}
\operatorname{Tr}\left\{\mathbf{MN}\right\}=\sum_{i=1}^{M_\mathrm{S}} \lambda_{i}(\mathbf{M N}) \leq \sum_{i=1}^{M_\mathrm{S}} \lambda_{i}(\mathbf{M}) \lambda_{i}(\mathbf{N}),
\label{a5}
\end{split}
\end{align}
where \( \lambda_i(\cdot) \) denotes the eigenvalues. They are arranged in descending order such that \( \lambda_{1}(\mathbf{M}) \geq \lambda_{2}(\mathbf{M}) \geq \cdots \geq \lambda_{M_\mathrm{S}}(\mathbf{M}) \) and \( \lambda_{1}(\mathbf{N}) \geq \lambda_{2}(\mathbf{N}) \geq \cdots \geq \lambda_{M_\mathrm{S}}(\mathbf{N}) \). From the \textit{Cauchy-Buniakowsky-Schwarz Inequality},
\begin{align}
	\begin{split}
& \sum_{i=1}^{M_\mathrm{S}} \lambda_{i}(\mathbf{M}) \lambda_{j}(\mathbf{N})\leq \sqrt{\left[\sum_{i=1}^{M_\mathrm{S}} {\lambda_{i}(\mathbf{M})}^2\right] \left[\sum_{j=1}^{M_\mathrm{S}}{\lambda_{j}(\mathbf{N})}^2\right]}
 \\& \leq \left[\sum_{i=1}^{M_\mathrm{S}} \lambda_{i}(\mathbf{M})\right]\left[\sum_{j=1}^{M_\mathrm{S}}\lambda_{j}(\mathbf{N})\right] =\operatorname{Tr}\left\{\mathbf{M}\right\}\operatorname{Tr}\left\{\mathbf{N}\right\},
		\label{a6}
	\end{split}
\end{align}
i.e.,
$\operatorname{Tr}\left\{\mathbf{MN}\right\}
\leq \operatorname{Tr}\left\{\mathbf{M}\right\}\operatorname{Tr}\left\{\mathbf{N}\right\}$. Similarly, we define $\mathbf{T}=\bar{\boldsymbol{\Upsilon}}_{\mathrm{ss}}\mathbf{A}^{-2}-\alpha$ and  $\mathbf{S}=\boldsymbol{\Upsilon}^{\mathrm {int}}_{\mathrm{sg}}$ and we have $\operatorname{Tr}\left\{\mathbf{TS}\right\}
\leq \operatorname{Tr}\left\{\mathbf{T}\right\}\operatorname{Tr}\left\{\mathbf{S}\right\}$. Substituting these inequalities into \eqref{a4},
\begin{equation}
\begin{split}
\nabla_{\varsigma } I_\mathrm{sg}
\leq& \operatorname{Tr}\left\{\frac{2\beta^2}{\alpha}\bar{\boldsymbol{\Upsilon}}_{\mathrm{ss}}\mathbf{A}^{-1}\boldsymbol{\Upsilon}^{\mathrm {int}}_{\mathrm{sg}}\mathbf{A}^{-2}\right\}\\&\times \operatorname{Tr}\left\{\mathbf{A}^{-1}\boldsymbol{\Upsilon}^{\mathrm {int}}_{\mathrm{sg}} \bar{\boldsymbol{\Upsilon}}_{\mathrm{ss}}\mathbf{A}^{-1}-\alpha\boldsymbol{\Upsilon}^{\mathrm {int}}_{\mathrm{sg}} \right\}
 \leq  0.
\label{a9}
\end{split}
\end{equation}
This concludes the proof.

	\ifCLASSOPTIONcaptionsoff
	\newpage
	\fi
	\bibliographystyle{IEEEtran}
	\bibliography{IEEEfull}

\begin{thebibliography}{10}
\providecommand{\url}[1]{#1}
\csname url@samestyle\endcsname
\providecommand{\newblock}{\relax}
\providecommand{\bibinfo}[2]{#2}
\providecommand{\BIBentrySTDinterwordspacing}{\spaceskip=0pt\relax}
\providecommand{\BIBentryALTinterwordstretchfactor}{4}
\providecommand{\BIBentryALTinterwordspacing}{\spaceskip=\fontdimen2\font plus
\BIBentryALTinterwordstretchfactor\fontdimen3\font minus
  \fontdimen4\font\relax}
\providecommand{\BIBforeignlanguage}[2]{{%
\expandafter\ifx\csname l@#1\endcsname\relax
\typeout{** WARNING: IEEEtran.bst: No hyphenation pattern has been}%
\typeout{** loaded for the language `#1'. Using the pattern for}%
\typeout{** the default language instead.}%
\else
\language=\csname l@#1\endcsname
\fi
#2}}
\providecommand{\BIBdecl}{\relax}
\BIBdecl

\bibitem{8088533}
L.~Kuang, X.~Chen, C.~Jiang, H.~Zhang, and S.~Wu, ``Radio resource management
  in future terrestrial-satellite communication networks,'' \emph{IEEE Wirel.
  Commun.}, vol.~24, no.~5, pp. 81--87, Oct. 2017.

\bibitem{8746876}
A.~I. Perez-Neira, M.~A. Vazquez, M.~B. Shankar, S.~Maleki, and S.~Chatzinotas,
  ``Signal processing for high-throughput satellites: Challenges in new
  interference-limited scenarios,'' \emph{IEEE Signal Process. Mag.}, vol.~36,
  no.~4, pp. 112--131, Jul. 2019.

\bibitem{8403528}
G.~Hattab, P.~Moorut, E.~Visotsky, M.~Cudak, and A.~Ghosh, ``Interference
  analysis of the coexistence of {5G} cellular networks with satellite earth
  stations in 3.7-4.2{GHz},'' in \emph{Proc. IEEE Int. Conf. Commun. Workshops
  (ICC Workshops)}, Jul. 2018, pp. 1--6.

\bibitem{5586925}
V.~Deslandes, J.~Tronc, and A.-L. Beylot, ``Analysis of interference issues in
  integrated satellite and terrestrial mobile systems,'' in \emph{Proc. Adv.
  Satell. Multimedia Syst. Conf. Signal Process. Space Commun. Workshop
  (ASMS/SPSC)}, 2010, pp. 256--261.

\bibitem{7811844}
M.~Jia, X.~Gu, Q.~Guo, W.~Xiang, and N.~Zhang, ``Broadband hybrid
  satellite-terrestrial communication systems based on cognitive radio toward
  {5G},'' \emph{IEEE Wirel. Commun.}, vol.~23, no.~6, pp. 96--106, Dec. 2016.

\bibitem{9052737}
E.~Lagunas, C.~G. Tsinos, S.~K. Sharma, and S.~Chatzinotas, ``{5G} cellular and
  fixed satellite service spectrum coexistence in {C}-band,'' \emph{IEEE
  Access}, vol.~8, pp. 72\,078--72\,094, 2020.

\bibitem{9385374}
X.~Fang, W.~Feng, T.~Wei, Y.~Chen, N.~Ge, and C.-X. Wang, ``{5G} embraces
  satellites for {6G} ubiquitous {IoT}: Basic models for integrated satellite
  terrestrial networks,'' \emph{IEEE Internet Things J.}, vol.~8, no.~18, pp.
  14\,399--14\,417, Sep. 2021.

\bibitem{9485040}
D.~Peng, A.~Bandi, Y.~Li, S.~Chatzinotas, and B.~Ottersten, ``Hybrid
  beamforming, user scheduling, and resource allocation for integrated
  terrestrial-satellite communication,'' \emph{IEEE Trans. Veh. Technol.},
  vol.~70, no.~9, pp. 8868--8882, Sep. 2021.

\bibitem{10645749}
H.~Nguyen-Kha, V.~Nguyen~Ha, E.~Lagunas, S.~Chatzinotas, and J.~Grotz, ``Joint
  two-tier user association and resource management for integrated
  satellite-terrestrial networks,'' \emph{IEEE Trans. Wireless Commun.},
  vol.~23, no.~11, pp. 16\,648--16\,665, Nov. 2024.

\bibitem{10301688}
H.-W. Lee, C.-C. Chen, C.-I.~S. Liao, A.~Medles, D.~Lin, I.-K. Fu, and H.-Y.
  Wei, ``Interference mitigation for reverse spectrum sharing in {B5G}/{6G}
  satellite-terrestrial networks,'' \emph{IEEE Trans. Veh. Technol.}, vol.~73,
  no.~3, pp. 4247--4263, Mar. 2024.

\bibitem{9749196}
D.~Peng, D.~He, Y.~Li, and Z.~Wang, ``Integrating terrestrial and satellite
  multibeam systems toward {6G}: Techniques and challenges for interference
  mitigation,'' \emph{IEEE Wirel. Commun.}, vol.~29, no.~1, pp. 24--31, Feb.
  2022.

\bibitem{9344715}
T.~Wei, W.~Feng, Y.~Chen, C.-X. Wang, N.~Ge, and J.~Lu, ``Hybrid
  satellite-terrestrial communication networks for the maritime internet of
  things: Key technologies, opportunities, and challenges,'' \emph{IEEE
  Internet Things J.}, vol.~8, no.~11, pp. 8910--8934, Jun. 2021.

\bibitem{7336495}
E.~Lagunas, S.~K. Sharma, S.~Maleki, S.~Chatzinotas, and B.~Ottersten,
  ``Resource allocation for cognitive satellite communications with incumbent
  terrestrial networks,'' \emph{IEEE Trans. Cogn. Commun. Netw.}, vol.~1,
  no.~3, pp. 305--317, Sep. 2015.

\bibitem{lagunas2015power}
{Lagunas, Eva and Sharma, Shree Krishna and Maleki, Sina and Chatzinotas,
  Symeon and Ottersten, Bjorn}, ``Power control for satellite uplink and
  terrestrial fixed-service co-existence in {Ka}-band,'' in \emph{Proc. IEEE
  Veh. Technol. Conf. (VTC2015-Fall)}, Sep. 2015, pp. 1--5.

\bibitem{sat1207}
G.~Ziaragkas, G.~Poziopoulou, J.~Núñez-Martínez, J.~Baranda, I.~Moreno,
  C.~Tsinos, S.~Maleki, S.~K. Sharma, M.~Alodeh, and S.~Chatzinotas,
  ``{SANSA}—hybrid terrestrial–satellite backhaul network: scenarios, use
  cases, {KPIs}, architecture, network and physical layer techniques,''
  \emph{Int. J. Satell. Commun. Networking}, vol.~35, no.~5, pp. 379--405, Oct.
  2017.

\bibitem{lagunas2017carrier}
E.~Lagunas, S.~Maleki, L.~Lei, C.~Tsinos, S.~Chatzinotas, and B.~Ottersten,
  ``Carrier allocation for hybrid satellite-terrestrial backhaul networks,'' in
  \emph{IEEE Int. Conf. Commun. Workshops (ICC Workshops)}, Jun. 2017, pp.
  718--723.

\bibitem{7986310}
C.~Jiang, X.~Zhu, L.~Kuang, Y.~Qian, and J.~Lu, ``Multimedia multicast
  beamforming in integrated terrestrial-satellite networks,'' in \emph{IEEE
  Int. Wirel. Commun. Mob. Comput. Conf., (IWCMC)}, Jun. 2017, pp. 340--345.

\bibitem{7559751}
H.~Baek and J.~Lim, ``Spectrum sharing for coexistence of fixed satellite
  services and frequency hopping tactical data link,'' \emph{IEEE J. Sel. Areas
  Commun.}, vol.~34, no.~10, pp. 2642--2649, Oct. 2016.

\bibitem{7387774}
K.~An, M.~Lin, W.-P. Zhu, Y.~Huang, and G.~Zheng, ``Outage performance of
  cognitive hybrid satellite-terrestrial networks with interference
  constraint,'' \emph{IEEE Trans. Veh. Technol.}, vol.~65, no.~11, pp.
  9397--9404, Nov. 2016.

\bibitem{9110855}
L.~You, K.-X. Li, J.~Wang, X.~Gao, X.-G. Xia, and B.~Ottersten, ``Massive
  {MIMO} transmission for {LEO} satellite communications,'' \emph{IEEE J. Sel.
  Areas Commun.}, vol.~38, no.~8, pp. 1851--1865, Aug. 2020.

\bibitem{Pastukh2023ChallengesOU}
A.~Pastukh, V.~Tikhvinskiy, S.~S. Dymkova, and O.~V. Varlamov, ``Challenges of
  using the {L}-band and {S}-band for direct-to-cellular satellite {5G-6G NTN}
  systems,'' \emph{Technologies}, 2023.

\bibitem{10179219}
J.~Heo, S.~Sung, H.~Lee, I.~Hwang, and D.~Hong, ``{MIMO} satellite
  communication systems: A survey from the {PHY} layer perspective,''
  \emph{IEEE Commun. Surv. Tutor.}, vol.~25, no.~3, pp. 1543--1570, 2023.

\bibitem{7887756}
V.~Joroughi, M.~Ã. Vázquez, A.~I. Pérez-Neira, and B.~Devillers, ``Onboard
  beam generation for multibeam satellite systems,'' \emph{IEEE Trans. Wireless
  Commun.}, vol.~16, no.~6, pp. 3714--3726, Jun. 2017.

\bibitem{10256078}
M.~Khammassi, A.~Kammoun, and M.-S. Alouini, ``Precoding for high-throughput
  satellite communication systems: A survey,'' \emph{IEEE Commun. Surv.
  Tutor.}, vol.~26, no.~1, pp. 80--118, 2024.

\bibitem{9992172}
G.~Geraci, D.~López-Pérez, M.~Benzaghta, and S.~Chatzinotas, ``Integrating
  terrestrial and non-terrestrial networks: {3D} opportunities and
  challenges,'' \emph{IEEE Commun. Mag.}, vol.~61, no.~4, pp. 42--48, Apr.
  2023.

\bibitem{10820534}
K.~Ntontin, E.~Lagunas, J.~Querol, J.~u. Rehman, J.~Grotz, S.~Chatzinotas, and
  B.~Ottersten, ``A vision, survey, and roadmap toward space communications in
  the {6G} and beyond era,'' \emph{Proc. IEEE}, pp. 1--37, 2025.

\bibitem{10068542}
D.~Tuzi, T.~Delamotte, and A.~Knopp, ``Satellite swarm-based antenna arrays for
  {6G} direct-to-cell connectivity,'' \emph{IEEE Access}, vol.~11, pp.
  36\,907--36\,928, 2023.

\bibitem{1494410}
T.~Yoo and A.~Goldsmith, ``Optimality of zero-forcing beamforming with
  multiuser diversity,'' in \emph{Proc. IEEE Int. Conf. Commun. (ICC)}, vol.~1,
  May 2005, pp. 542--546.

\bibitem{1391204}
C.~Peel, B.~Hochwald, and A.~Swindlehurst, ``A vector-perturbation technique
  for near-capacity multiantenna multiuser communication-part {I}: Channel
  inversion and regularization,'' \emph{IEEE Trans. Commun.}, vol.~53, no.~1,
  pp. 195--202, Jan. 2005.

\bibitem{4712693}
S.~S. Christensen, R.~Agarwal, E.~De~Carvalho, and J.~M. Cioffi, ``Weighted
  sum-rate maximization using weighted {MMSE} for {MIMO}-{BC} beamforming
  design,'' \emph{IEEE Trans. Wireless Commun.}, vol.~7, no.~12, pp.
  4792--4799, Dec. 2008.

\bibitem{wang2024}
\BIBentryALTinterwordspacing
Y.~Wang, H.~Hou, X.~Yi, W.~Wang, and S.~Jin, ``Towards unified {AI} models for
  {MU-MIMO} communications: A tensor equivariance framework,'' \emph{arXiv
  preprint arXiv:2406.09022}, 2024. [Online]. Available:
  \url{https://arxiv.org/abs/2406.09022}
\BIBentrySTDinterwordspacing

\bibitem{10437228}
S.~Wu, Y.~Wang, G.~Sun, L.~You, W.~Wang, and R.~Ding, ``Energy and
  computational efficient precoding for {LEO} satellite communications,'' in
  \emph{Proc. IEEE Glob. Commun. Conf (GLOBECOM)}, Dec. 2023, pp. 1872--1877.

\bibitem{6832894}
E.~Björnson, M.~Bengtsson, and B.~Ottersten, ``Optimal multiuser transmit
  beamforming: A difficult problem with a simple solution structure,''
  \emph{IEEE Signal Process. Mag.}, vol.~31, no.~4, pp. 142--148, Jul. 2014.

\bibitem{9815078}
Y.~Liu, Y.~Wang, J.~Wang, L.~You, W.~Wang, and X.~Gao, ``Robust downlink
  precoding for {LEO} satellite systems with per-antenna power constraints,''
  \emph{IEEE Trans. Veh. Technol.}, vol.~71, no.~10, pp. 10\,694--10\,711, Oct.
  2022.

\bibitem{10038746}
H.~Dong, C.~Hua, L.~Liu, W.~Xu, S.~Guo, and R.~Tafazolli, ``Joint beamformer
  design and user scheduling for integrated terrestrial-satellite networks,''
  \emph{IEEE Trans. Wireless Commun.}, vol.~22, no.~10, pp. 6398--6414, Oct.
  2023.

\bibitem{10.1049/iet-com.2018.5313}
Y.~Jiang, J.~Ouyang, C.~Yin, Z.~Xu, X.~Tao, and L.~Lou, ``Downlink beamforming
  scheme for hybrid satellite-terrestrial networks,'' \emph{IET Commun.},
  vol.~12, no.~18, p. 2342–2346, Oct. 2018.

\bibitem{6636830}
S.~K. Sharma, S.~Chatzinotas, and B.~Ottersten, ``Transmit beamforming for
  spectral coexistence of satellite and terrestrial networks,'' in \emph{Proc.
  IEEE Int. Conf. Cognitive Radio Oriented Wirel. Networks Commun.,
  (CROWNCOM)}, Jul. 2013, pp. 275--281.

\bibitem{9502017}
Q.~Wang, H.~Zhang, J.-B. Wang, F.~Yang, and G.~Y. Li, ``Joint beamforming for
  integrated mmwave satellite-terrestrial self-backhauled networks,''
  \emph{IEEE Trans. Veh. Technol.}, vol.~70, no.~9, pp. 9103--9117, Sep. 2021.

\bibitem{8886590}
Y.~Zhang, L.~Yin, C.~Jiang, and Y.~Qian, ``Joint beamforming design and
  resource allocation for terrestrial-satellite cooperation system,''
  \emph{IEEE Trans. Commun.}, vol.~68, no.~2, pp. 778--791, Feb. 2020.

\bibitem{8933099}
H.~Zhang, C.~Jiang, J.~Wang, L.~Wang, Y.~Ren, and L.~Hanzo, ``Multicast
  beamforming optimization in cloud-based heterogeneous terrestrial and
  satellite networks,'' \emph{IEEE Trans. Veh. Technol.}, vol.~69, no.~2, pp.
  1766--1776, Feb. 2020.

\bibitem{8894065}
C.~Liu, W.~Feng, Y.~Chen, C.-X. Wang, and N.~Ge, ``Optimal beamforming for
  hybrid satellite terrestrial networks with nonlinear {PA} and imperfect
  {CSIT},'' \emph{IEEE Wireless Commun. Lett.}, vol.~9, no.~3, pp. 276--280,
  Mar. 2020.

\bibitem{8292945}
M.~A. Vázquez, L.~Blanco, and A.~I. Pérez-Neira, ``Hybrid analog digital
  transmit beamforming for spectrum sharing backhaul networks,'' \emph{IEEE
  Trans. Signal Process.}, vol.~66, no.~9, pp. 2273--2285, May 2018.

\bibitem{8353853}
X.~Zhu, C.~Jiang, L.~Yin, L.~Kuang, N.~Ge, and J.~Lu, ``Cooperative multigroup
  multicast transmission in integrated terrestrial-satellite networks,''
  \emph{IEEE J. Sel. Areas Commun.}, vol.~36, no.~5, pp. 981--992, May 2018.

\bibitem{20251017998213}
D.~Kim, S.~Cho, W.~Shin, J.~Park, and D.~K. Kim, ``Distributed precoding for
  satellite-terrestrial integrated networks without sharing {CSIT}: A
  rate-splitting approach,'' \emph{IEEE Trans. Wireless Commun.}, 2025.

\bibitem{3GPPTR38.811}
3GPP, ``Study on new radio ({NR}) to support non-terrestrial networks,'' 3rd
  Generation Partnership Project, Tech. Rep. TR 38.811, Sep. 2020.

\bibitem{iet12629}
S.~Wu, G.~Sun, Y.~Wang, L.~You, W.~Wang, and R.~Ding, ``Low-complexity user
  scheduling for {LEO} satellite communications,'' \emph{IET Commun.}, vol.~17,
  no.~12, pp. 1368--1383, May 2023.

\bibitem{9427230}
C.~Wu, X.~Yi, Y.~Zhu, W.~Wang, L.~You, and X.~Gao, ``Channel prediction in
  high-mobility massive {MIMO}: From spatio-temporal autoregression to deep
  learning,'' \emph{IEEE J. Sel. Areas Commun.}, vol.~39, no.~7, pp.
  1915--1930, Jul. 2021.

\bibitem{4655459}
N.~Letzepis and A.~J. Grant, ``Capacity of the multiple spot beam satellite
  channel with rician fading,'' \emph{IEEE Trans. Inf. Theory}, vol.~54,
  no.~11, pp. 5210--5222, Nov. 2008.

\bibitem{3GPPTR38.863}
3GPP, ``Solutions for {NR} to support non-terrestrial networks ({NTN}):
  Non-terrestrial networks ({NTN}) related {RF} and co-existence aspects,'' 3rd
  Generation Partnership Project, Tech. Rep. TR 38.863, Mar. 2025.

\bibitem{fcc_sat_loa}
{Federal Communications Commission (FCC)}, ``{Satellite Space Stations:
  Application to Launch and Operate},'' FCC.report.
  \url{https://fcc.report/IBFS/Filing-List/SAT-LOA} (accessed: Apr. 17, 2025).

\bibitem{8795582}
W.~Wang, Y.~Tong, L.~Li, A.-A. Lu, L.~You, and X.~Gao, ``Near optimal timing
  and frequency offset estimation for {5G} integrated {LEO} satellite
  communication system,'' \emph{IEEE Access}, vol.~7, pp. 113\,298--113\,310,
  2019.

\bibitem{10107609}
J.~Shi, A.-A. Lu, W.~Zhong, X.~Gao, and G.~Y. Li, ``Robust {WMMSE} precoder
  with deep learning design for massive {MIMO},'' \emph{IEEE Trans. Commun.},
  vol.~71, no.~7, pp. 3963--3976, Jul. 2023.

\bibitem{8314727}
K.~Shen and W.~Yu, ``Fractional programming for communication systems—part
  {I}: Power control and beamforming,'' \emph{IEEE Trans. Signal Process.},
  vol.~66, no.~10, pp. 2616--2630, May 2018.

\bibitem{1570864}
M.~Villalobos and Y.~Zhang, ``A trust-region interior-point method for
  nonlinear programming,'' in \emph{TAPIA - Richard Tapia Celebr. Diversity
  Comput. Conf.}, Oct. 2005, pp. 7--9.

\bibitem{10440321}
Z.~Xiang, X.~Gao, K.-X. Li, and X.-G. Xia, ``Massive {MIMO} downlink
  transmission for multiple {LEO} satellite communication,'' \emph{IEEE Trans.
  Commun.}, vol.~72, no.~6, pp. 3352--3364, Jun. 2024.

\bibitem{5756489}
Q.~Shi, M.~Razaviyayn, Z.-Q. Luo, and C.~He, ``An iteratively weighted {MMSE}
  approach to distributed sum-utility maximization for a {MIMO} interfering
  broadcast channel,'' \emph{IEEE Trans. Signal Process.}, vol.~59, no.~9, pp.
  4331--4340, Sep. 2011.

\bibitem{10066300}
K.-X. Li, X.~Gao, and X.-G. Xia, ``Channel estimation for {LEO} satellite
  massive {MIMO} {OFDM} communications,'' \emph{IEEE Trans. Wireless Commun.},
  vol.~22, no.~11, pp. 7537--7550, Nov. 2023.

\bibitem{hou2024}
\BIBentryALTinterwordspacing
H.~Hou, Y.~Wang, Y.~Zhu, X.~Yi, W.~Wang, D.~T.~M. Slock, and S.~Jin, ``A
  tensor-structured approach to dynamic channel prediction for massive {MIMO}
  systems with temporal non-stationarity,'' \emph{arXiv preprint
  arXiv:2412.06713}, 2024. [Online]. Available:
  \url{https://arxiv.org/abs/2412.06713}
\BIBentrySTDinterwordspacing

\bibitem{Wang2013PenaltyFP}
K.-J. Wang and X.~Zhang, ``Penalty function-based precoding for downlink
  multiuser {MIMO} systems,'' \emph{AEU Int. J. Electron. Commun.}, vol.~67,
  pp. 167--173, Feb. 2013.

\end{thebibliography}

\end{document}